\documentclass[aps,prb,preprint,showpacs,superscriptaddress]{revtex4-1}
\usepackage{amsmath, amssymb, graphicx, bm}

\usepackage{siunitx}
\usepackage{graphicx}
\usepackage{float} 
\usepackage{tabularx}
\usepackage{threeparttable}
\usepackage{caption}
\newlength\myindention

\usepackage[utf8]{inputenc}
\usepackage[english]{babel}
\addto\captionsenglish{}
\usepackage{array}%
\usepackage{amsmath}
\usepackage{dcolumn}
\usepackage{url}
\usepackage{varioref}
\usepackage{subcaption}

\usepackage[colorlinks,urlcolor=blue,linkcolor=blue,citecolor=blue]{hyperref} 
\usepackage{multirow}

\newcommand{\is}[1]{_\text{\textit{#1}}}
\newcommand{\iu}[1]{^\text{\textit{#1}}}
\DeclareRobustCommand{\rchi}{{\mathpalette\irchi\relax}}
\newcommand{\irchi}[2]{\raisebox{\depth}{$#1\chi$}}
\newcommand{\comment}[1]{}
\begin{document}
\captionsetup[figure]{labelformat={default},labelsep=period,name={Figure.}}

\title{Influence of the presence of multiple resonances on material parameter determination using broadband ferromagnetic resonance spectroscopy}

\author{Prabandha Nakarmi} 
 \affiliation{Department of Physics and Astronomy, The University of Alabama, Tuscaloosa, Alabama $35487$, USA} 

\author{Tim Mewes}
\email[Corresponding author:]{tmewes@ua.edu}
\affiliation{Department of Physics and Astronomy, The University of Alabama, Tuscaloosa, Alabama $35487$, USA}

\date{\today}

\begin{abstract}
\textnormal
The influence of the presence of multiple resonances in ferromagnetic resonance spectra on extracted material parameters is investigated using numerical simulations. Our results show that the systematic error of assuming an incorrect number of resonances for a material can lead to the extraction of material parameters that significantly deviate from any of the true material parameters. When noise is present in experimental ferromagnetic resonance spectra increasing the frequency range of the broadband characterization can significantly reduce the error-margins when the data is analyzed assuming the correct number of resonances present in the material. For the cases investigated in this study it was found that analyzing the data using a single resonance results in extracted gyromagnetic ratios and effective magnetization parameters that are consistent with the weighted average of the true material parameters. We further provide a cautionary example regarding the extraction of the inhomogeneous linewidth broadening and damping parameters of materials that contain an unknown number of resonances.
\end{abstract}

\maketitle

\setlength{\parindent}{5ex}

\section{\label{sec:level1}Introduction}
\begin{figure}
\centering
\includegraphics[height = 4in, width = 6in, keepaspectratio]{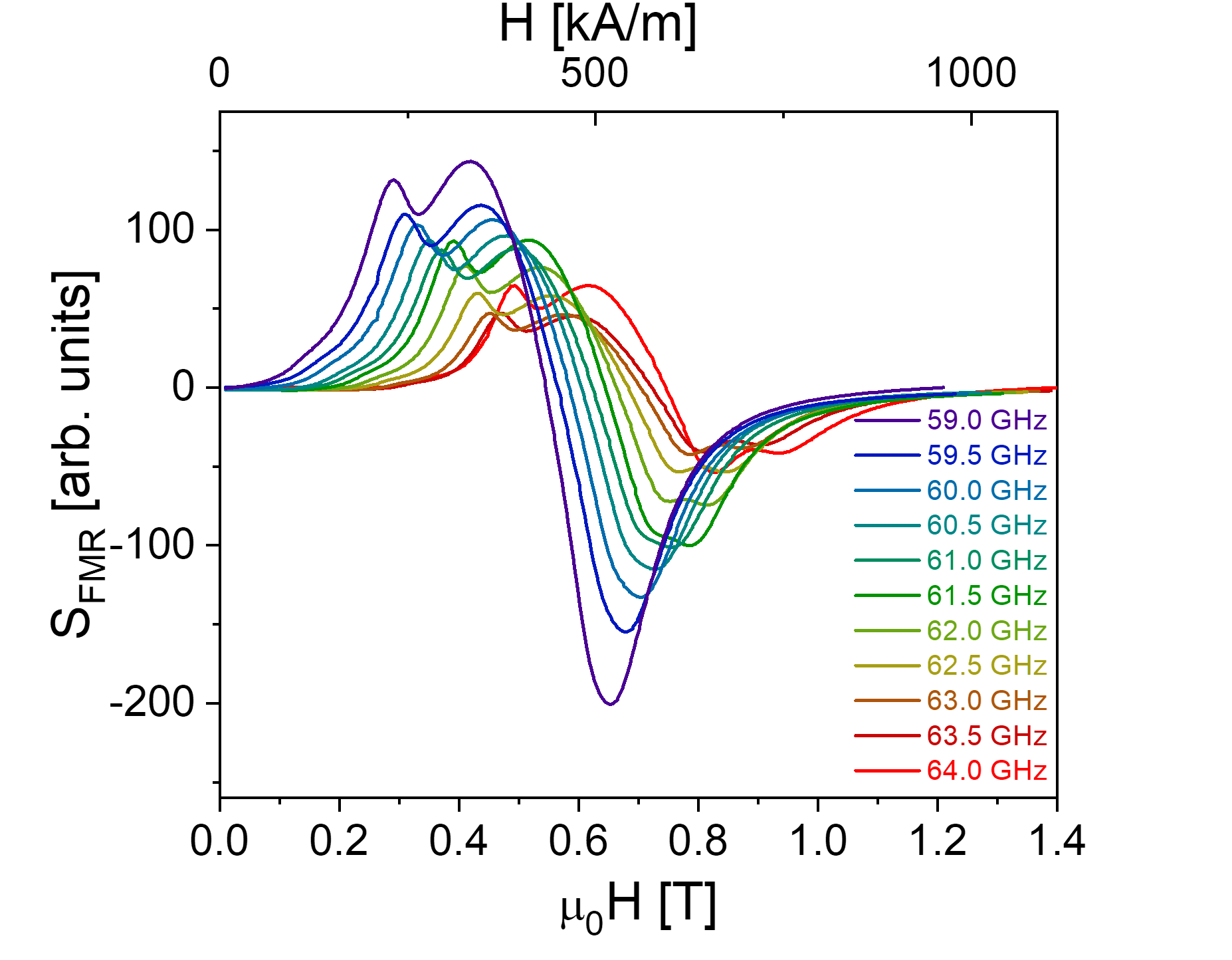}
\caption{Experimental ferromagnetic resonance spectra for an M-type hexaferrite \cite{Qorvo} measured with the magnetic field applied along the out-of-plane direction over a microwave frequency range from from $59\, [GHz]$ to $64\, [GHz]$ with spectra recorded in $0.5\, [GHz]$ intervals. } 
\label{fig:experimentalResonances}
\end{figure}

Ferromagnetic resonance spectroscopy (FMR) is a well established characterization method for magnetic materials \cite{Arkadyev1912,VonSovskii2016,Farle1998,Mewes2021}. While early experimental techniques were based on resonant cavities operated at a single frequency, recent advancements have led to broadband capabilities based on the use of coplanar waveguides \cite{Sugiyama1993,Zhang1997,bilzer2007,Bilzer2007b,Lee2009,Schafer2012}, vector network analyzers \cite{Weir1974,Counil2004,Ding2004,Kuanr2005,Neudecker2006,Kalarickal2006,bilzer2007,Godsell2010}, and electrically detected FMR \cite{tsoi2000,kiselev2003,tulapurkar2005,Gui2005,Sankey2006,Harder2011,wang2018}. Independent of the experimental details used to carry out these measurements ferromagnetic resonance spectroscopy can provide important information about magnetic material properties including the gyromagnetic ratio $\gamma'$ and the effective magnetization $M\is{eff}$ of the sample under investigation.
This is based on the condition for ferromagnetic resonance in a magnetic material that links the resonance frequency $f$ with the free energy density $e$ of the system, which according to Smit and Beljers \cite{smit1955} can be written as:
\cite{smit1955,Baselgia1988}:
\begin{equation}
    \left(\frac{f}{\gamma_0'}\right)^2=\frac{1}{M_s^2 \sin^2 \theta}\left[\frac{\partial^2e}{\partial\theta^2}\frac{\partial^2e}{\partial\phi^2}-\left(\frac{\partial^2e}{\partial\theta\partial\phi}\right)^2\right],
    \label{eq:smit-beljers}
\end{equation} 
where $\gamma_0'=\mu_0|\gamma'|$ is the reduced gyromagnetic ratio $\gamma'=\frac{\gamma}{2\pi}$ rescaled by the vacuum permeability $\mu_0$, $M_s$ is the saturation magnetization, $\theta$ the polar angle, and $\phi$ the azimuthal angle of the magnetization. The general expression for the resonance condition given by equation \eqref{eq:smit-beljers} can be simplified when the sample is saturated along a high symmetry direction. In the case of a thin film one can show that if the sample is saturated along the out-of-plane direction that the resonance condition is given by \cite{Kittel1947,Kittel1948}:
\begin{equation}
    f=\gamma_0'\left(H-M\is{eff}\right)\label{eq:Kittel}.
\end{equation}
Here we have introduced the effective magnetization $M\is{eff}=M_s-\frac{2K_u}{\mu_0 M_S}$ that takes into account the potential presence of a perpendicular anisotropy $K_u$. The perpendicular anisotropy has the same functional dependence on the polar angle $\theta$ as the shape anisotropy and therefore cannot be distinguished from it using only FMR. Equation \eqref{eq:Kittel} shows that in this particular measurement geometry the resonance frequency of the system depends linearly on the applied field with the slope given by the gyromagnetic ratio, just like in electron spin resonance (ESR)\cite{Poole1967}. The only modification compared to ESR is the presence of the internal field characterized by $M\is{eff}$. Furthermore, equation \eqref{eq:Kittel} clearly shows that in this situation for a given microwave frequency $f$ the resonance condition is met for exactly one field value, the resonance field $H=H_{res}$.\\
However, throughout the history of ferromagnetic resonance spectroscopy there have been numerous examples in the literature showing multiple resonances
\cite{Shin1987,Talisa1988,Aslam2020,Artman1991,Omaggio1979,Cofield1986,Vukadinovic2001,Pechan2005,Pechan2006,Zhang1991,Goennenwein2003}. The reasons given in the literature for the existence of multiple resonances are as diverse as the samples that have been characterized using FMR. They include for example compositional variations \cite{Shin1987,Talisa1988,Aslam2020}, different anisotropies \cite{Artman1991}, anisotropy variation across the film thickness \cite{Omaggio1979,Cofield1986}, unsaturated samples \cite{Vukadinovic2001}, edge modes \cite{Pechan2005,Pechan2006}, vortex modes \cite{Pechan2006}, coupling of multilayers \cite{Zhang1991}, and spin wave resonances \cite{Goennenwein2003}. \\
Therefore when faced with multiple resonances in FMR experiments it can be challenging to pinpoint the origin of those resonances or to predict how many resonances are expected for a particular sample. Figure \ref{fig:experimentalResonances} shows an example of experimental spectra featuring multiple resonances in an M-type hexaferrite sample \cite{Qorvo} measured using broadband ferromagnetic resonance spectroscopy \cite{Mewes2021}.
\\
In order to obtain insights regarding the influence of assumptions made during the data analysis process of spectra featuring an unknown number of resonances, we will focus in this article on numerical simulations. For the simulations we assume the presence of multiple resonances in a material with a strong perpendicular anisotropy. All resonances are assumed to fulfill equation \eqref{eq:Kittel} albeit with slightly different material parameters. The simulated data is subsequently analyzed just like one would analyze experimental ferromagnetic resonance spectroscopy data. As the number of resonances that contribute to an experimental resonance spectrum is not known a priori, we investigate in detail the influence of assuming different numbers of resonances on the results of the data analysis. 
\section{Numerical techniques}
\begin{figure}
\centering
\includegraphics[height = 4in, width = 6in, keepaspectratio]{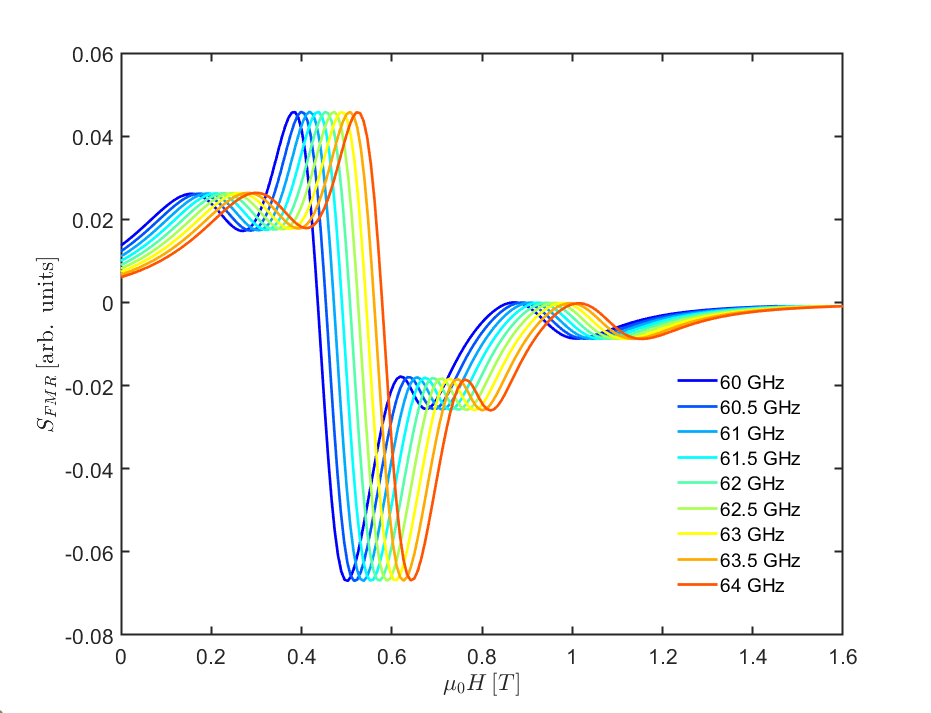}
\caption{Simulated field modulated FMR signal $S\is{FMR}(H)$  for a material that consists of four different constituents. Here $\gamma_{k}'=28\, [\frac{GHz}{T}]$, $\mu_0M\is{eff,k}=\{-1.9,-1.7,-1.5,-1.21\}\, [T]$, $\alpha_k=0.01$, $\mu_0\Delta H_{0,k}=\{200, 100, 75, 150\} [mT]$, $A_k=\{0.3, 1.0, 0.2, 0.1\}$, and $B_k=0$ for $k=\{1,...,4\}$. The spectra are calculated at the frequencies indicated in the legend using 200 field points each.} 
\label{fig:multipleResonances}
\end{figure}

Our simulations are aimed at reproducing data as it would be obtained experimentally in coplanar waveguide based broadband ferromagnetic resonance spectroscopy \cite{Sterwerf2016,Mewes2021}. In our simulations we assume that the material contains four different constituents with slightly different material properties. For each constituent $k$, with $k=\{1,...,4\}$, we define a separate gyromagnetic ratio $\gamma_{k}'$, effective magnetization $M\is{eff,k}$, Gilbert damping parameter $\alpha_k$, and inhomogeneous linewidth broadening $\Delta H_{0,k}$. 
For many broadband ferromagnetic resonance spectroscopy experimental realizations it is common to record spectra at a fixed frequency while sweeping the field. In addition, field modulation with lock-in detection is frequently used to improve the signal-to-noise ratio \cite{Rai2020,Klingler2014}. We therefore simulate field-swept spectra at different microwave frequencies $f$. The simulations assume a Lorentzian lineshape of the imaginary part of the magnetic susceptibility $\rchi''_k$ for all four constituents of the material. For sufficiently small field modulation the lock-in detection results in a recorded FMR signal proportional to the first derivative $\partial \rchi''_k/\partial H$ of the imaginary part of the susceptibility with respect to the field $H$. Generally the field dependent FMR signal $S\is{FMR,k}(H)$ contribution from constituent $k$ measured in a spectrometer can be written as follows \cite{Poole1967,Oates2002}:
\begin{equation}
S\is{FMR,k}(H)=\frac{A_k\left(\frac{H_{res,k}-H}{\Delta H_k/2}\right)+9B_k-3B_k\left(\frac{H_{res,k}-H}{\Delta H_k/2}\right)^2}{\left[3+\left(\frac{H_{res,k}-H}{\Delta H_k/2}\right)^2\right]^2},
\label{eq:lineshape}
\end{equation} 
where $A_k$ and $B_k$ are the amplitudes of the absorption and dispersion signal contributions to the measured signal. While ideally a spectrometer would only measure the absorption part of the signal, it is common to observe a mixture of both contributions experimentally \cite{Oates2002}. However, for simplicity we assume for the simulations that $B_k=0$ for $k=\{1,...,4\}$, i.e. we assume that there is no dispersion contribution to the FMR signals.
The other parameter that enters equation \eqref{eq:lineshape} is the frequency dependent peak-to-peak linewidth $\Delta H_k$, which we assume can also be different for each constituent $k$. In our simulations we assume that the peak-to-peak linewidth is caused by a Gilbert-like damping term in the Landau-Lifshitz-Gilbert equation of motion. In this case the frequency dependence of the linewidth is given by:
\begin{equation}
\Delta H_k(f)=\Delta H_{0,k}+\frac{2}{\sqrt{3}}\frac{\alpha_k}{\gamma_{0,k}'} f.
\label{eq:linewidth}
\end{equation}
This enables us to calculate the total signal $S\is{FMR}(H)$ for any choice of microwave frequency $f$ by summing the responses of the four different constituents:
\begin{equation}
S\is{FMR}(H)=\sum_{k=1}^{4}S\is{FMR,k}(H).
\label{eq:totallineshape}
\end{equation}
Figure \ref{fig:multipleResonances} shows an example of simulated FMR spectra using this methodology.
\\
This simulated broadband FMR data is then subjected to the same methodology one would typically use to fit experimental FMR data, see for example \cite{Klingler2014,Khodadadi2017,Wu2019}. As with experimental data the number of resonances or constituents that are present in the spectra is assumed to be unknown a priori. We therefore vary the number of resonances $N$ that we use when attempting to fit the simulated data. The fit function is thus the sum of $N$ resonances: 
\begin{equation}
S\iu{Fit}\is{FMR}(H)=\sum_{k=1}^{N}S\is{FMR,k}(H),
\label{eq:fit}
\end{equation}
where the $S\is{FMR,k}(H)$ are again given by equation \eqref{eq:lineshape}, but the resonance field, linewidth, absorption amplitude, and dispersion amplitude are now fit-parameters. We note here that, just like for experimental spectra, we do not assume for the fits that the dispersion amplitude is zero but instead keep this a free parameter of the fit. To distinguish the fit parameters from the original parameters used to simulate the spectra we will use a superscript, thus these parameters are labeled $H\iu{Fit}_{res,k}$, $\Delta H\iu{Fit}_k$, $A\iu{Fit}_k$, and $B\iu{Fit}_k$ respectively. \\
For each frequency $f$ the algorithm first estimates initial parameters to fit the spectrum using a single resonance. Once the fitting algorithm has converged to a solution, the residual between the data and the fitted curve is used to estimate the additional fit parameters for the fit with two resonances. Once this fit with two resonances converges, the process is repeated by including one more resonance, until a fit is obtained for the desired number of resonances $N$ of the fit. Our fitting algorithm takes advantage of global optimization tools available in Matlab \cite{MATLAB2020,MATLABGOTB}. This includes the option of using global search, multi start algorithms \cite{MatlabGS,Ugray2007}, and particle swarm optimization \cite{MatlabPS,Kennedy1995,Pedersen2010,Mezura2011}. However, empirically we have found that in most cases all three algorithms tend to lead to very similar results. Therefore we limit our analysis in this manuscript to using particle swarm optimization to search for the global minimum of the sum of squares $S=\sum_i{r_i^2}$ of the residuals $r_i=(S\is{FMR}(H_i)-S\iu{Fit}\is{FMR}(H_i))$.\\ 
Once the fit has converged for the desired $N$ resonances present in the fit function we compute $R^2$ and the adjusted-$R^2$ values \cite{Kvalseth1985}. Because we are interested in comparing fit attempts that use a different number of resonances $N$ the degrees of freedom in our fit functions will differ significantly. Therefore, we will limit our discussion to the adjusted-$R^2$ values of the fits. \\
To estimate the error-margins of the fit-parameters $H\iu{Fit}_{res,k}$, $\Delta H\iu{Fit}_k$, $A\iu{Fit}_k$, and $B\iu{Fit}_k$ we use bootstrapping \cite{Diciccio1996}. The resonance fields $H\iu{Fit}_{res,k}$ and the associated error-margins $\sigma_{H\iu{Fit}_{res,k}}$ with $k=\{1,...,N\}$ then serve as input for a linear fit using equation \eqref{eq:Kittel} to determine the gyromagnetic ratio ${\gamma\iu{Fit}_{k}}'$, effective magnetization $M\iu{Fit}\is{eff,k}$ and their respective error-margins.
\section{Results and Discussion}
In the following we will discuss three different cases of materials that all contain four constituents with different material properties. In section \ref{Meff-Variation} we analyze the case of a material with constituents that share the same gyromagnetic ratio but differ with respect to their effective magnetizations. This section also provides detailed examples of the methodology we used. In section \ref{gamma-variation} we analyze a material with constituents that differ with respect to their gyromagnetic ratios but share the same effective magnetization. Finally in section \ref{Noise} we revisit the case of a material with a shared gyromagnetic ratio and different effective magnetization but now analyze the influence of noise on the results. We conclude this section with a cautionary example regarding the extraction of damping related parameters using equation \eqref{eq:linewidth}.
\subsection{Constituents with different effective magnetization}\label{Meff-Variation}
\begin{figure}
     \centering
     \begin{subfigure}[b]{0.48\textwidth}
         \caption{$N=1$ \hfill\,}
         \centering
         \includegraphics[width=\textwidth,trim={0 0 0 0.65cm},clip]{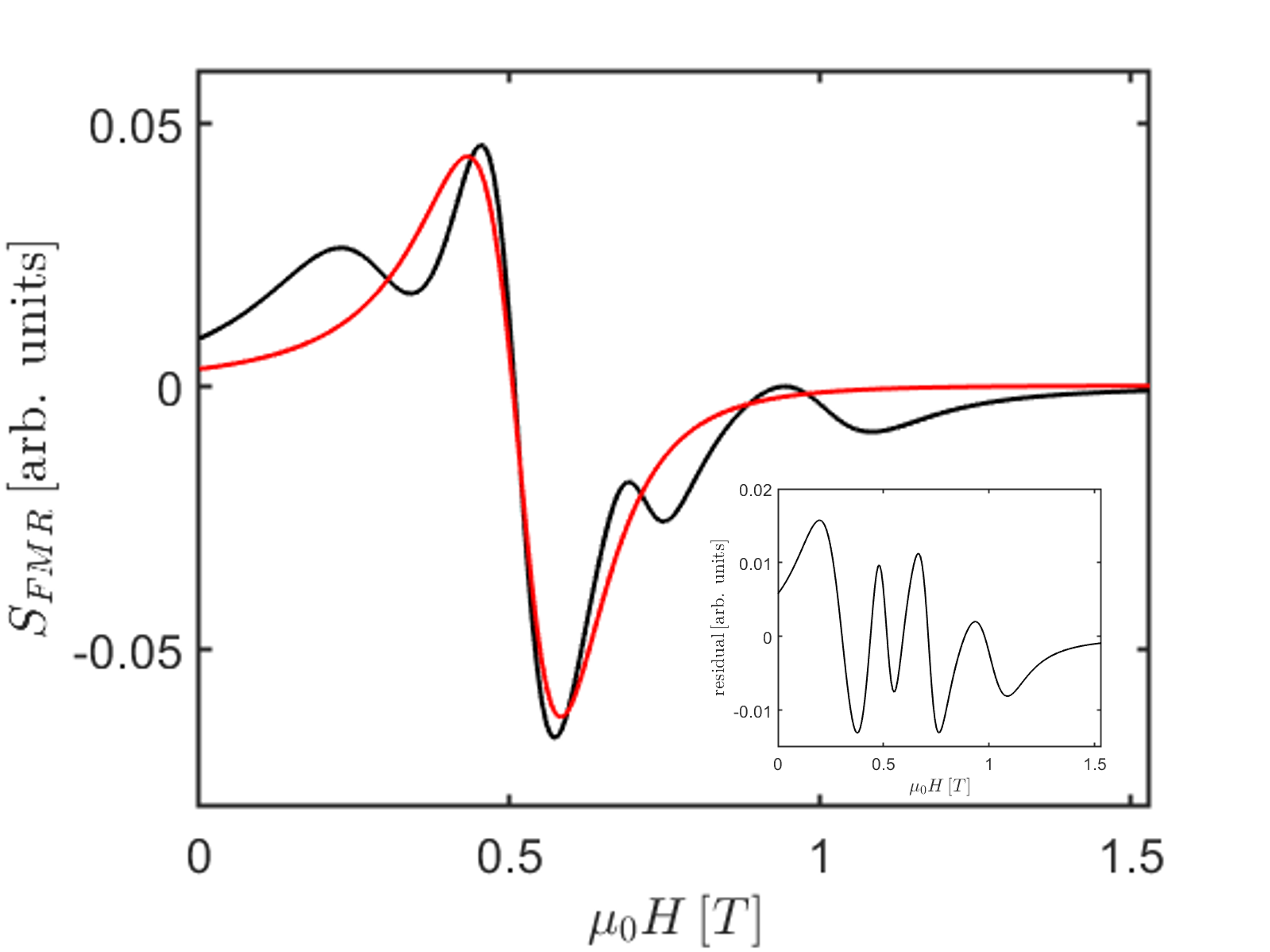}
     \end{subfigure}
     \begin{subfigure}[b]{0.48\textwidth}
         \centering
         \caption{$N=2$ \hfill\,}
         \includegraphics[width=\textwidth,trim={0 0 0 0.65cm},clip]{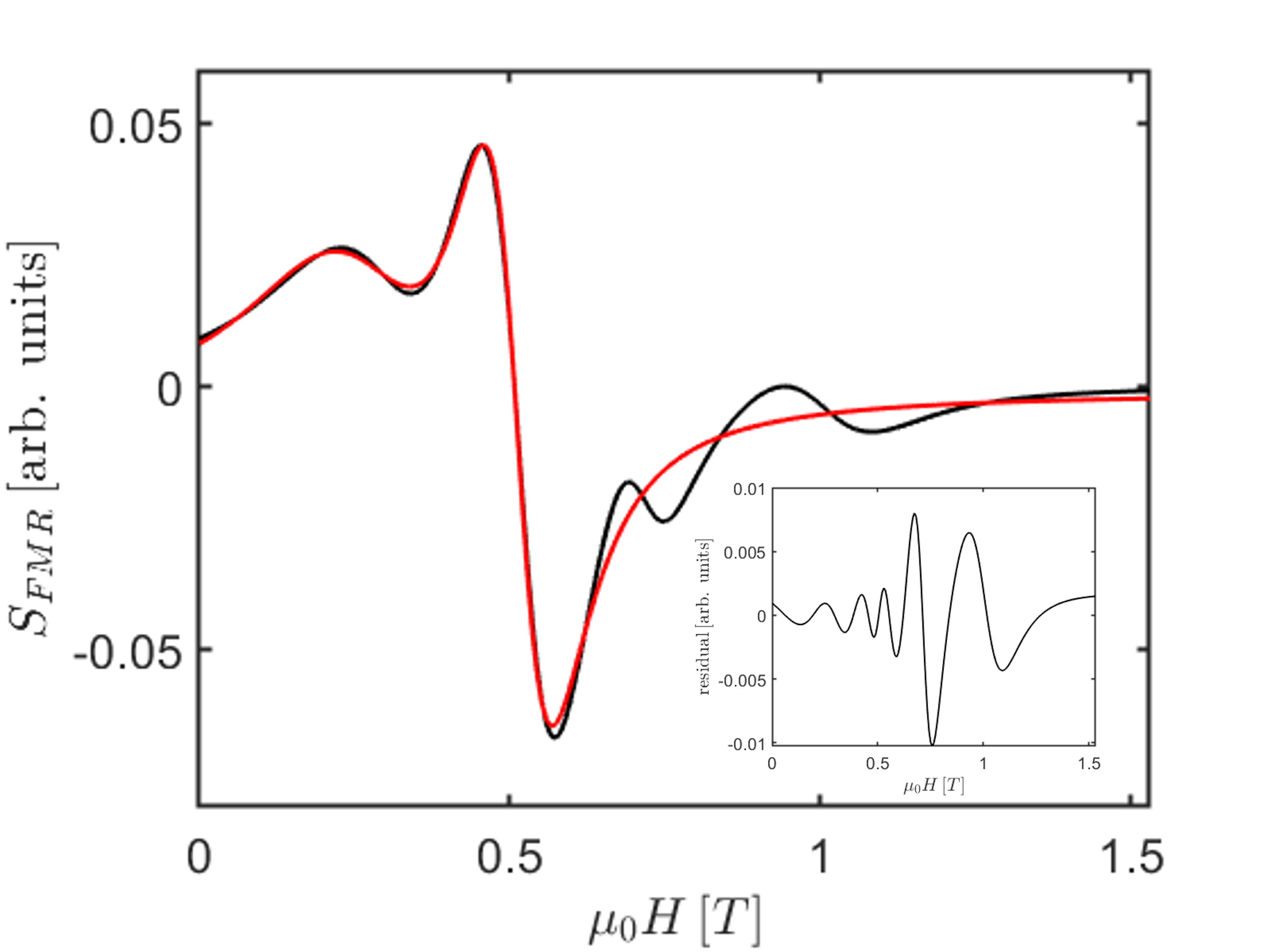}
     \end{subfigure}
     \begin{subfigure}[b]{0.48\textwidth}
         \caption{$N=3$ \hfill\,}
         \centering
         \includegraphics[width=\textwidth,trim={0 0 0 0.65cm},clip]{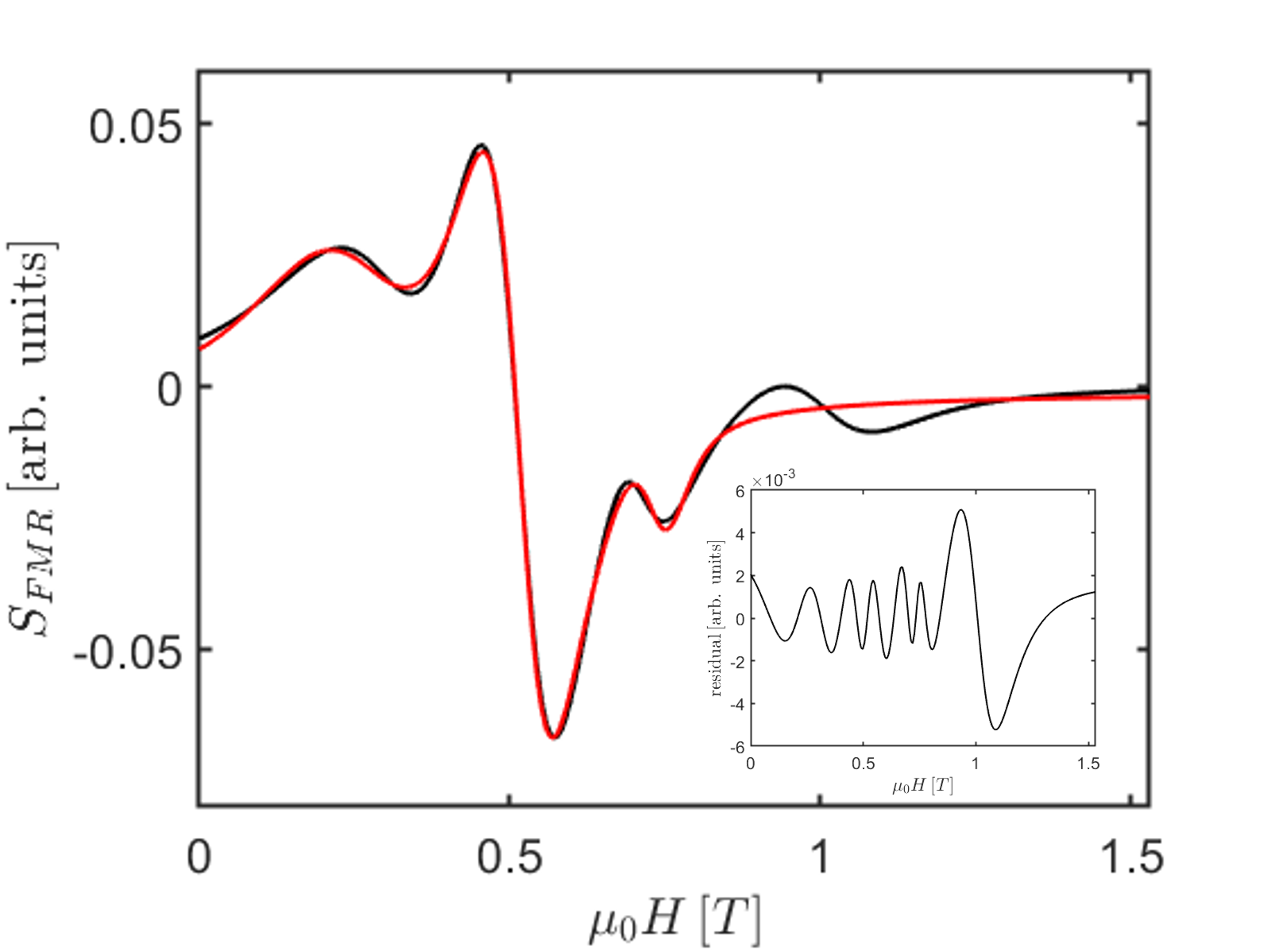}
     \end{subfigure}
     \begin{subfigure}[b]{0.48\textwidth}
         \centering
         \caption{$N=4$ \hfill\,}
         \includegraphics[width=\textwidth,trim={0 0 0 0.65cm},clip]{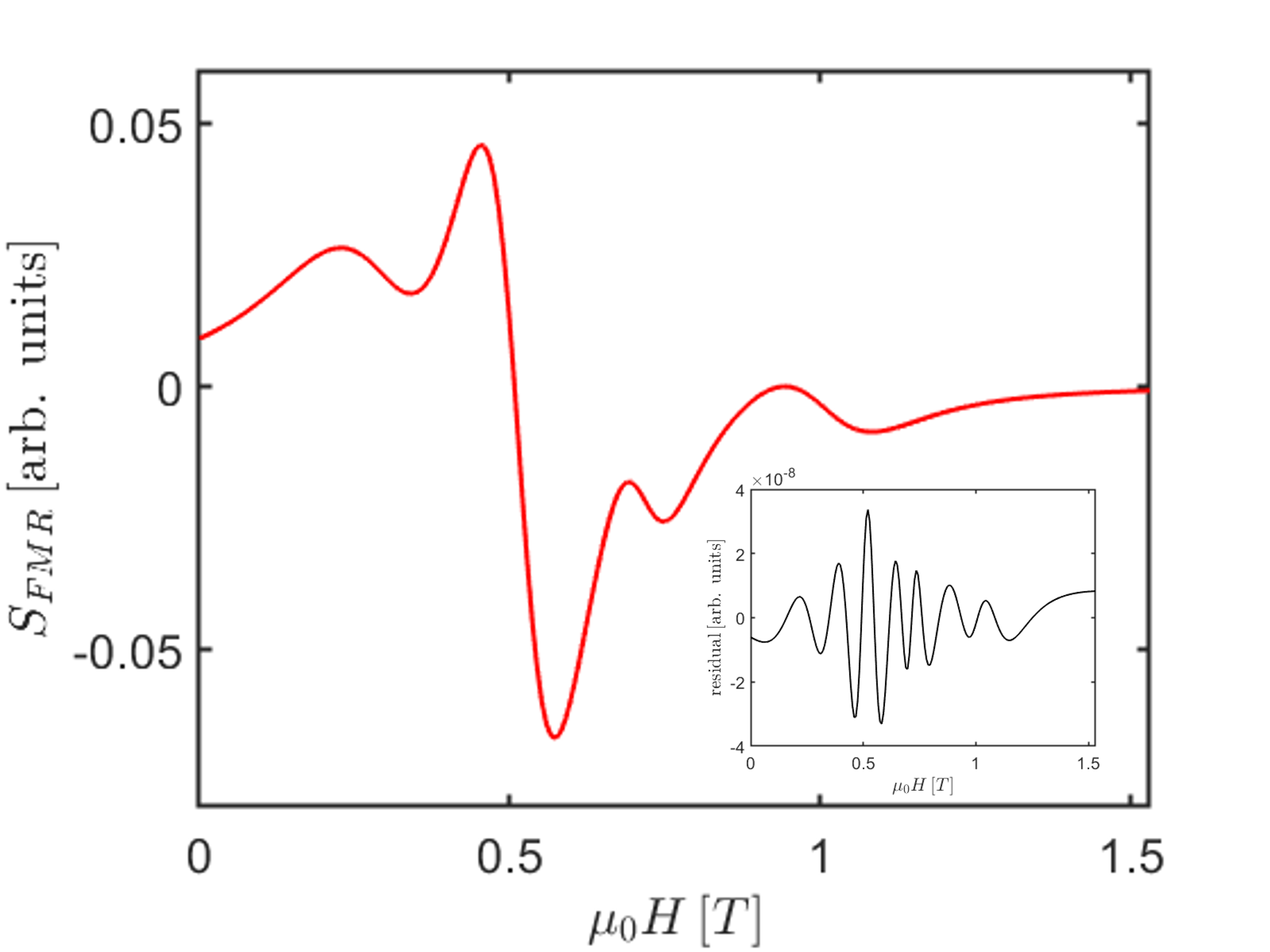}
     \end{subfigure}
     \begin{subfigure}[b]{0.48\textwidth}
         \caption{$N=5$ \hfill\,}
         \centering
         \includegraphics[width=\textwidth,trim={0 0 0 0.15cm},clip]{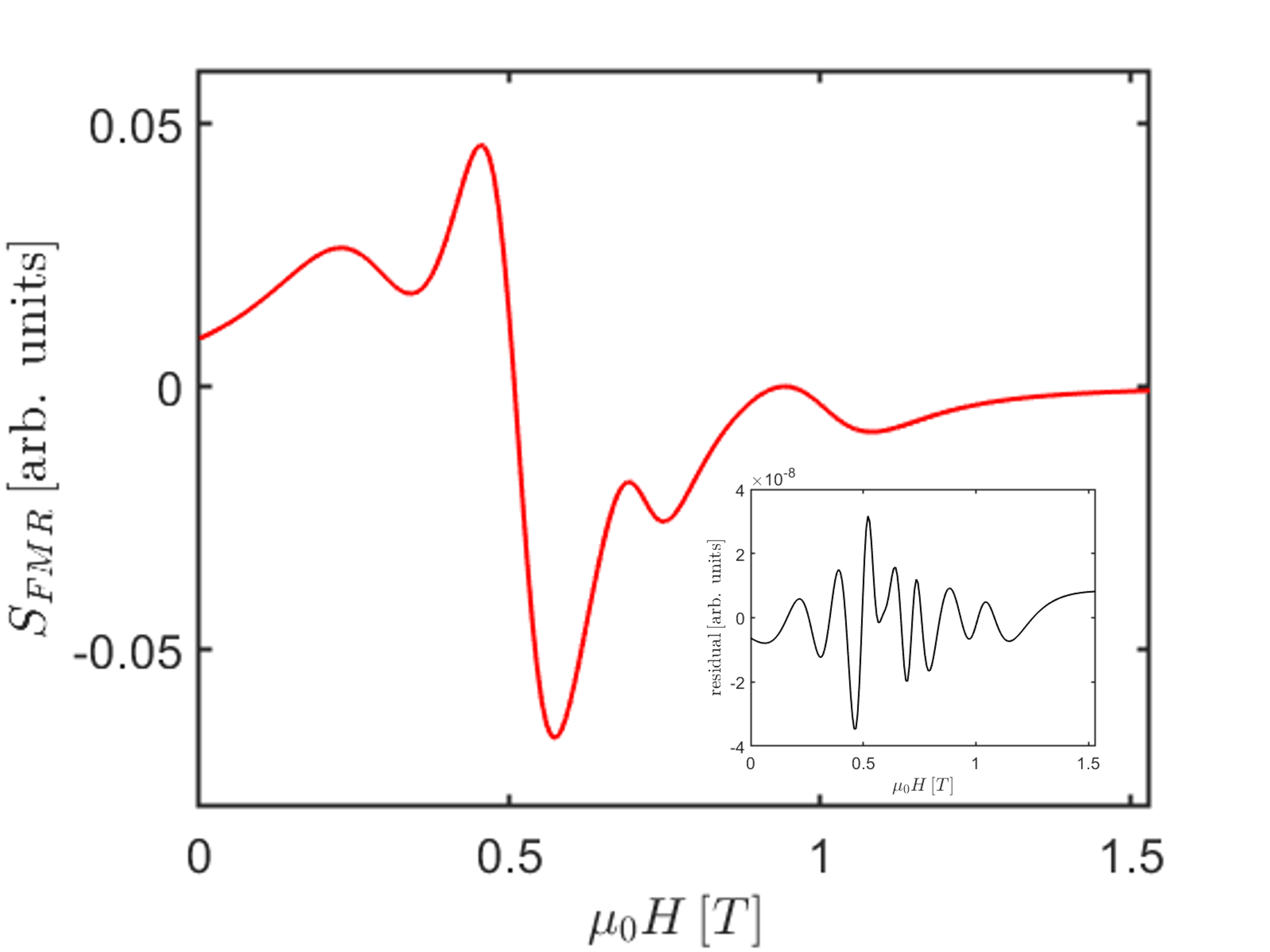}
     \end{subfigure}
     \begin{subfigure}[b]{0.48\textwidth}
         \caption{\hfill\,}
         \centering
     \end{subfigure}
        \caption{Exemplary spectra (black) and fits (red) for $f=62\,[GHz]$ and their corresponding residuals (insets). The fits (a)-(e) use an increasing number of resonances $N=\{1,...,5\}$ to fit the simulated spectrum that contains $k=4$ resonances.}
        \label{fig:ResonanceFits}
\end{figure}

\begin{figure}
     \centering
     \begin{subfigure}[b]{0.4\textwidth}
         \caption{$N=1$\hfill\,}
         \centering
         \includegraphics[width=\textwidth]{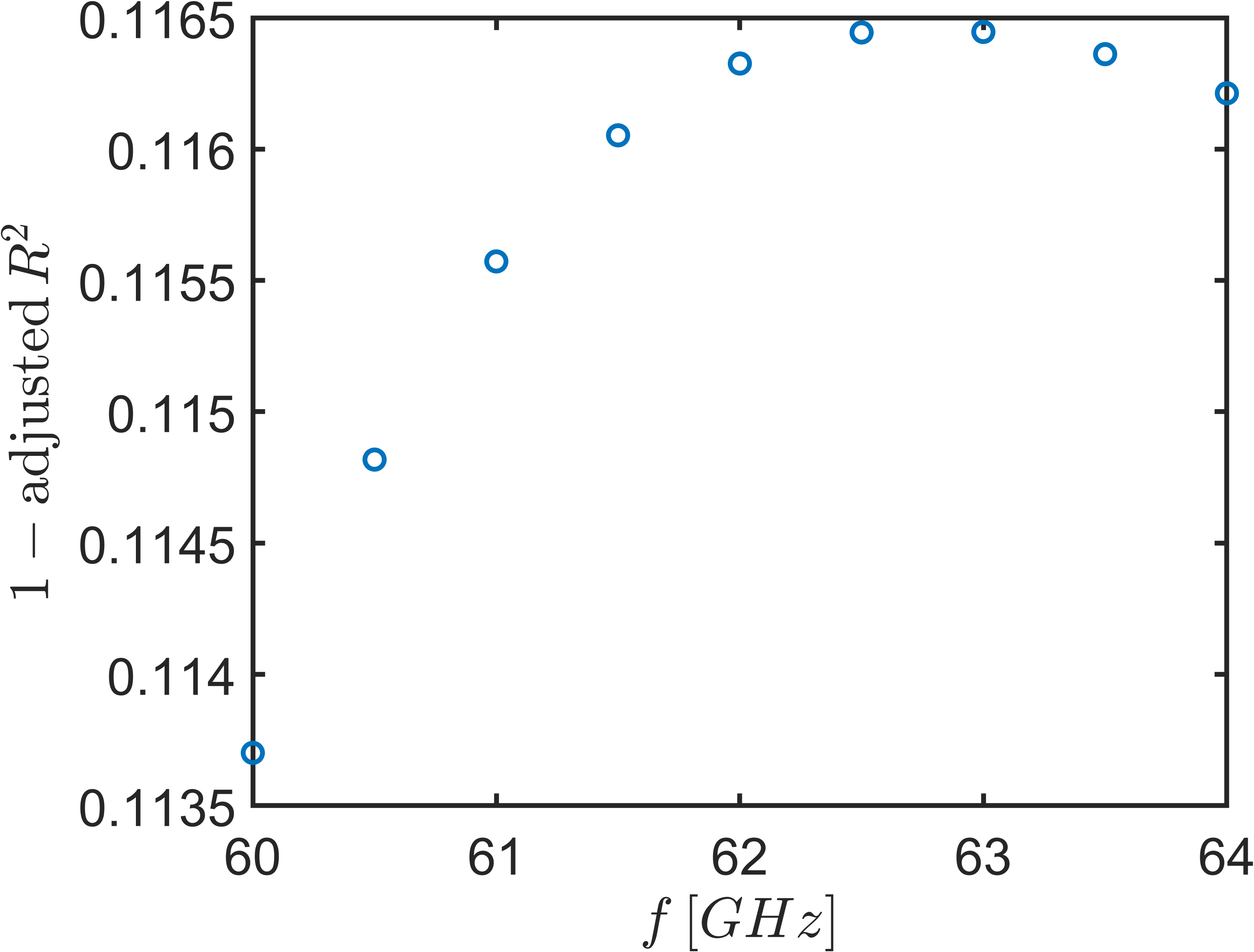}
     \end{subfigure}
     \hfill
     \begin{subfigure}[b]{0.4\textwidth}
         \centering
         \caption{$N=2$\hfill\,}
         \includegraphics[width=\textwidth]{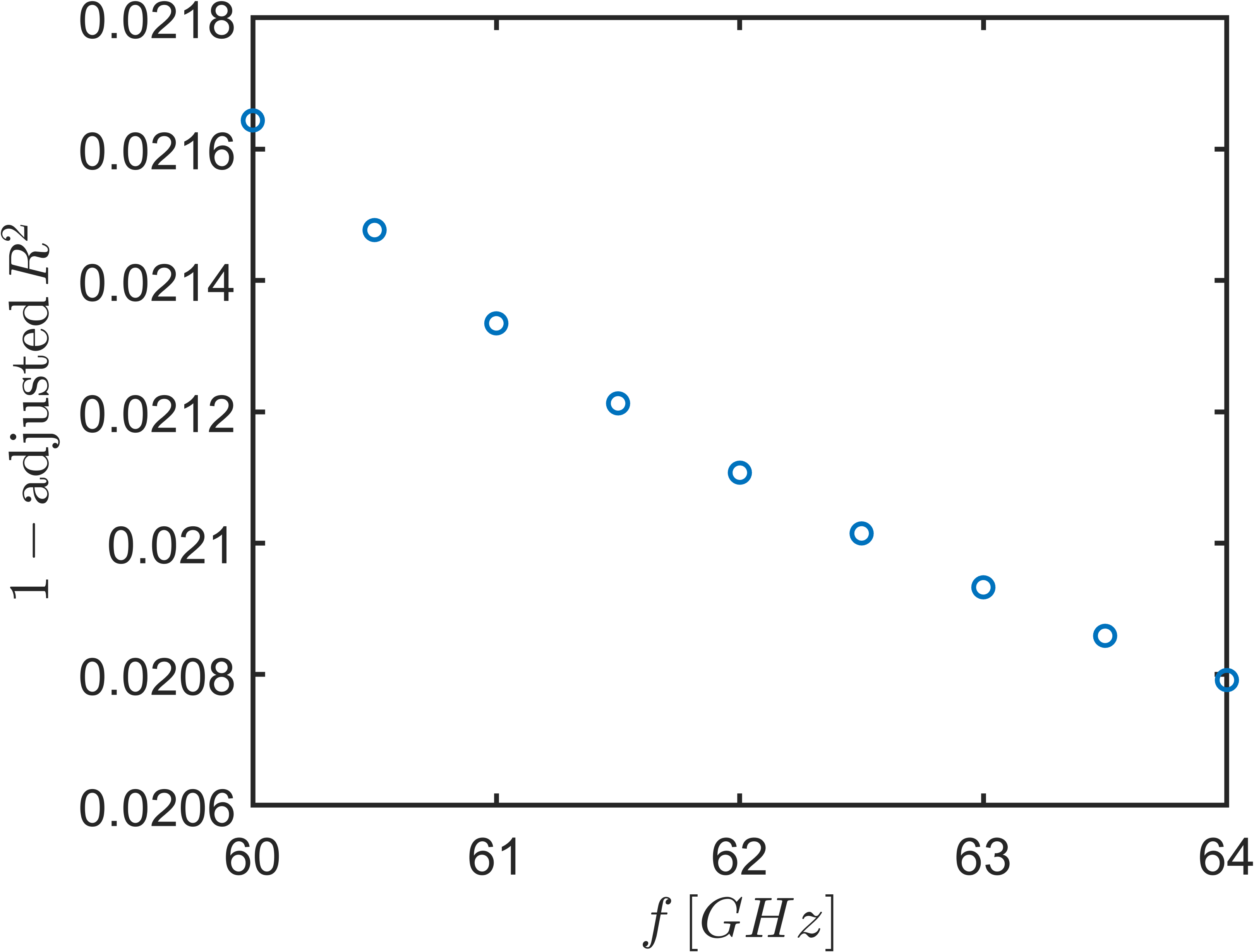}
     \end{subfigure}
     \begin{subfigure}[b]{0.4\textwidth}
         \caption{$N=3$\hfill\,}
         \centering
         \includegraphics[width=\textwidth,trim={-1.2cm 0 0 0}]{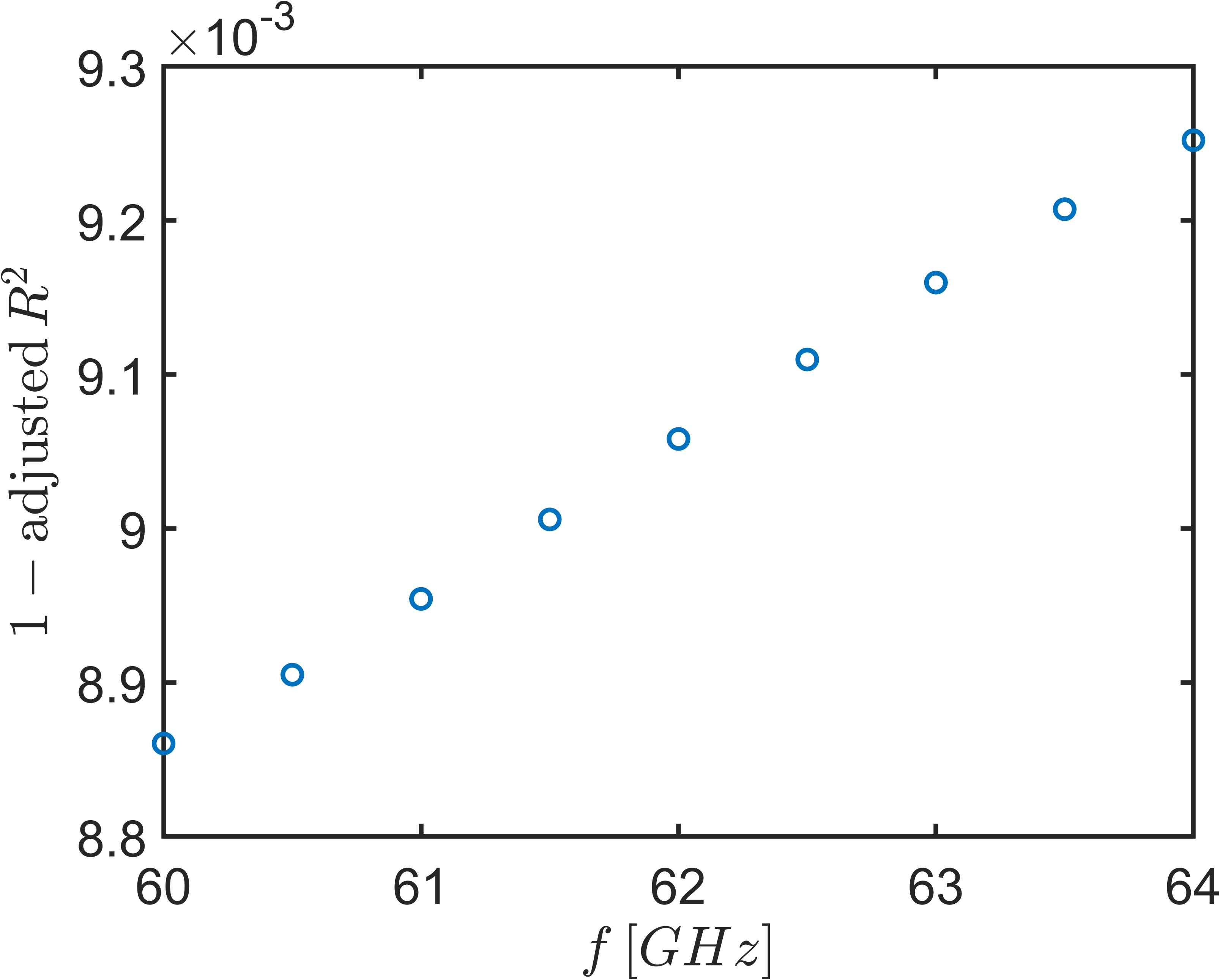}
     \end{subfigure}
     \hfill
     \begin{subfigure}[b]{0.4\textwidth}
         \centering
         \caption{$N=4$\hfill\,}
         \includegraphics[width=\textwidth,trim={-1.2cm 0 0 0}]{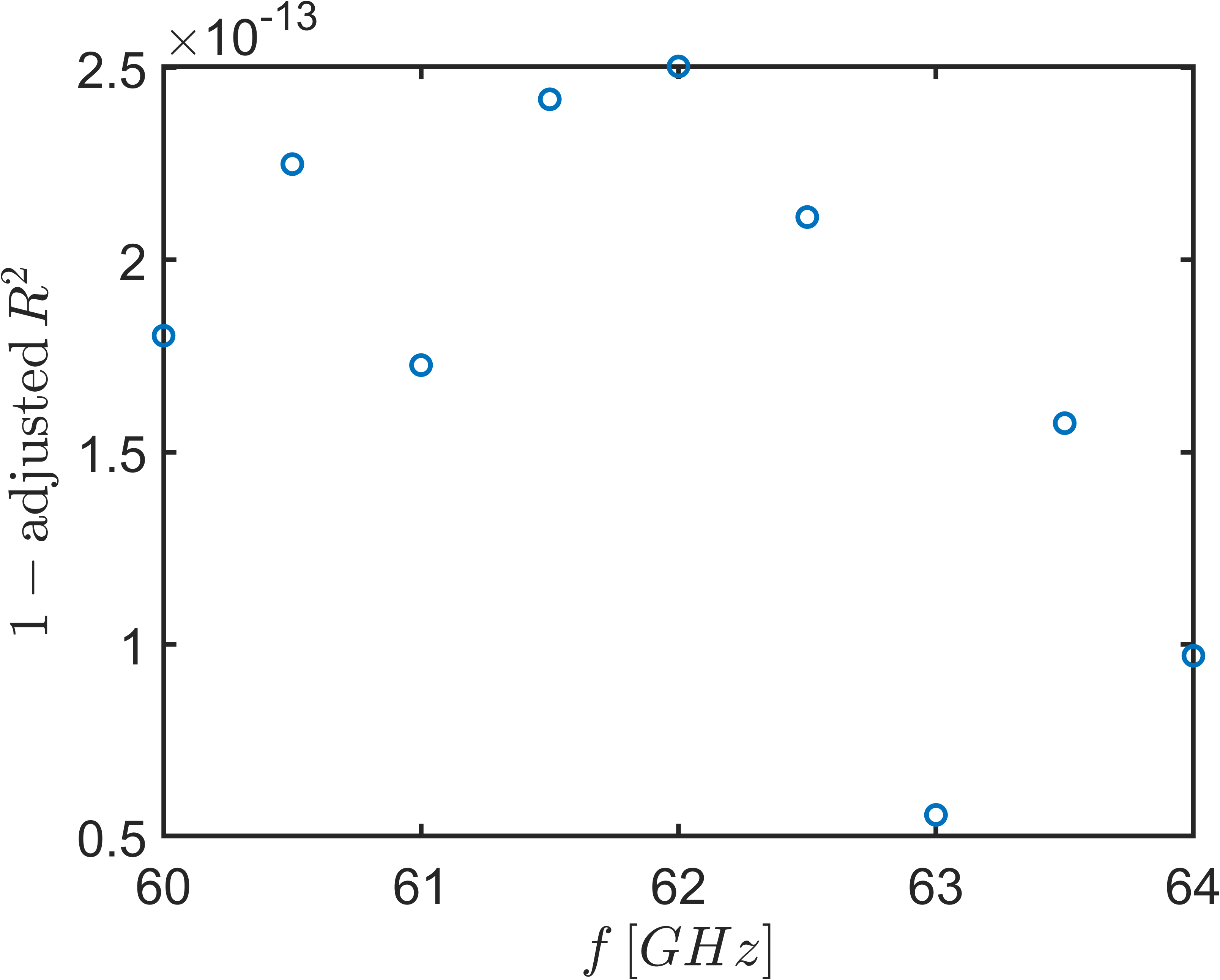}
     \end{subfigure}
        \caption{The deviation $(1-\text{adjusted-}R^2)$ for the fits of the spectra is shown for different number of resonances $N=\{1,...,4\}$ used to fit the simulated spectrum that contains $k=4$ resonances.}
        \label{fig:adjustedR2}
\end{figure}

\begin{figure}
     \centering
     \begin{subfigure}[b]{0.4\textwidth}
         \caption{$N=1$\hfill\,}
         \centering
         \includegraphics[width=\textwidth,trim={0.9cm 0 2.5cm 0},clip]{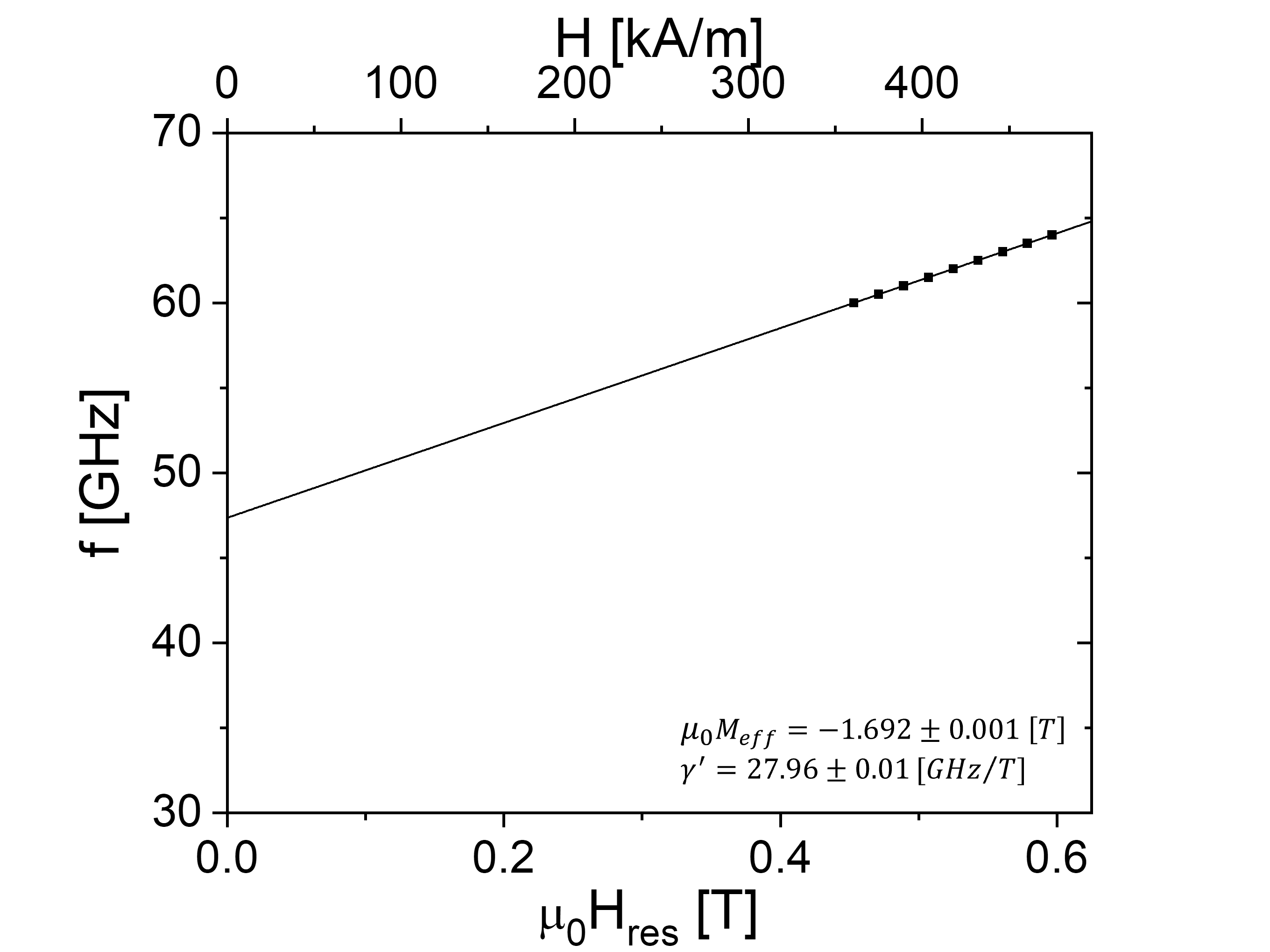}
     \end{subfigure}
     \hfill
     \begin{subfigure}[b]{0.4\textwidth}
         \centering
         \caption{$N=2$\hfill\,}
         \includegraphics[width=\textwidth,trim={0.9cm 0 2.5cm 0},clip]{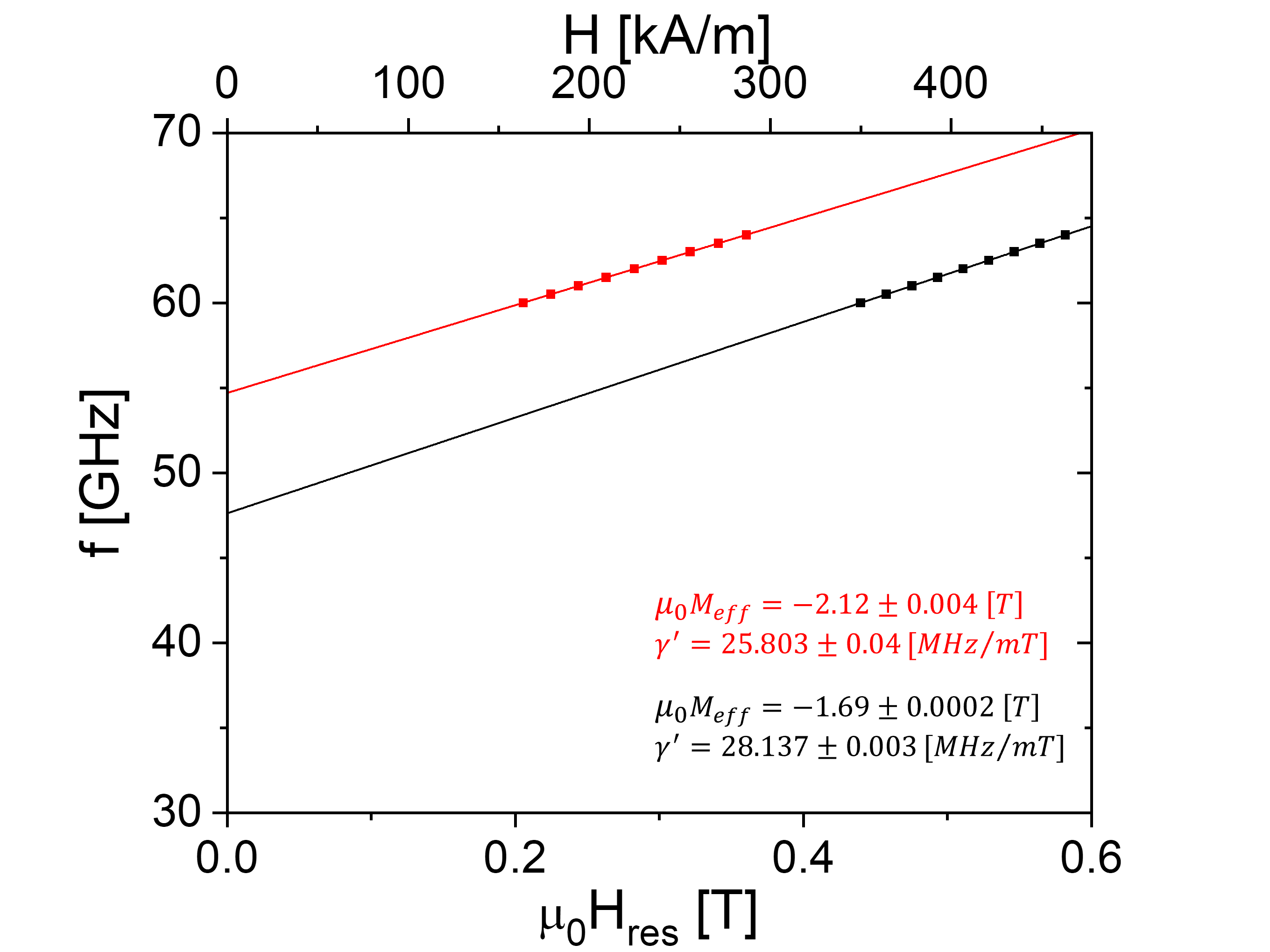}
     \end{subfigure}
     \begin{subfigure}[b]{0.4\textwidth}
         \caption{$N=3$\hfill\,}
         \centering
         \includegraphics[width=\textwidth,trim={0.9cm 0 2.5cm 0},clip]{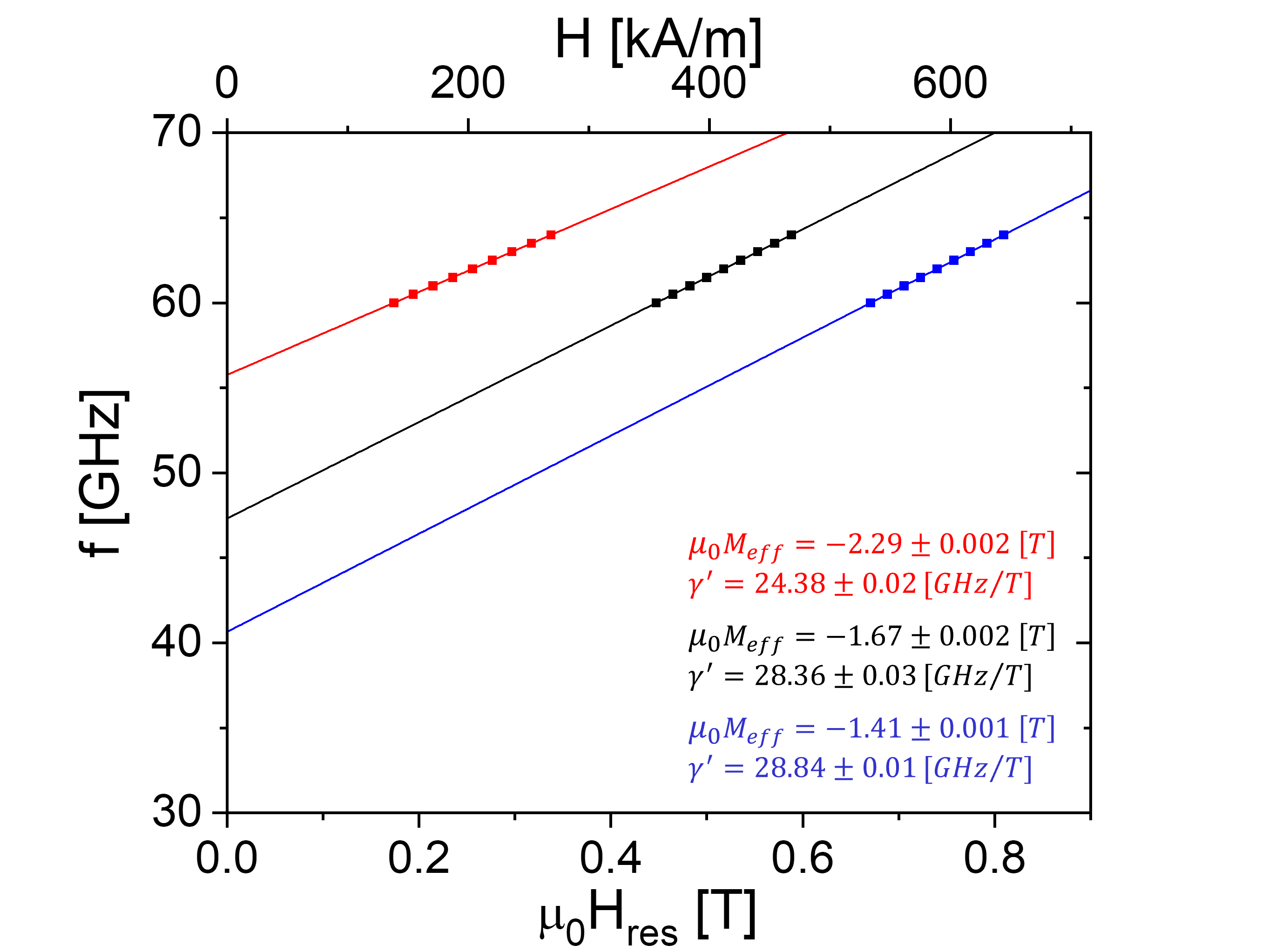}
     \end{subfigure}
     \hfill
     \begin{subfigure}[b]{0.4\textwidth}
         \centering
         \caption{$N=4$\hfill\,}
         \includegraphics[width=\textwidth,trim={0.9cm 0 2.5cm 0},clip]{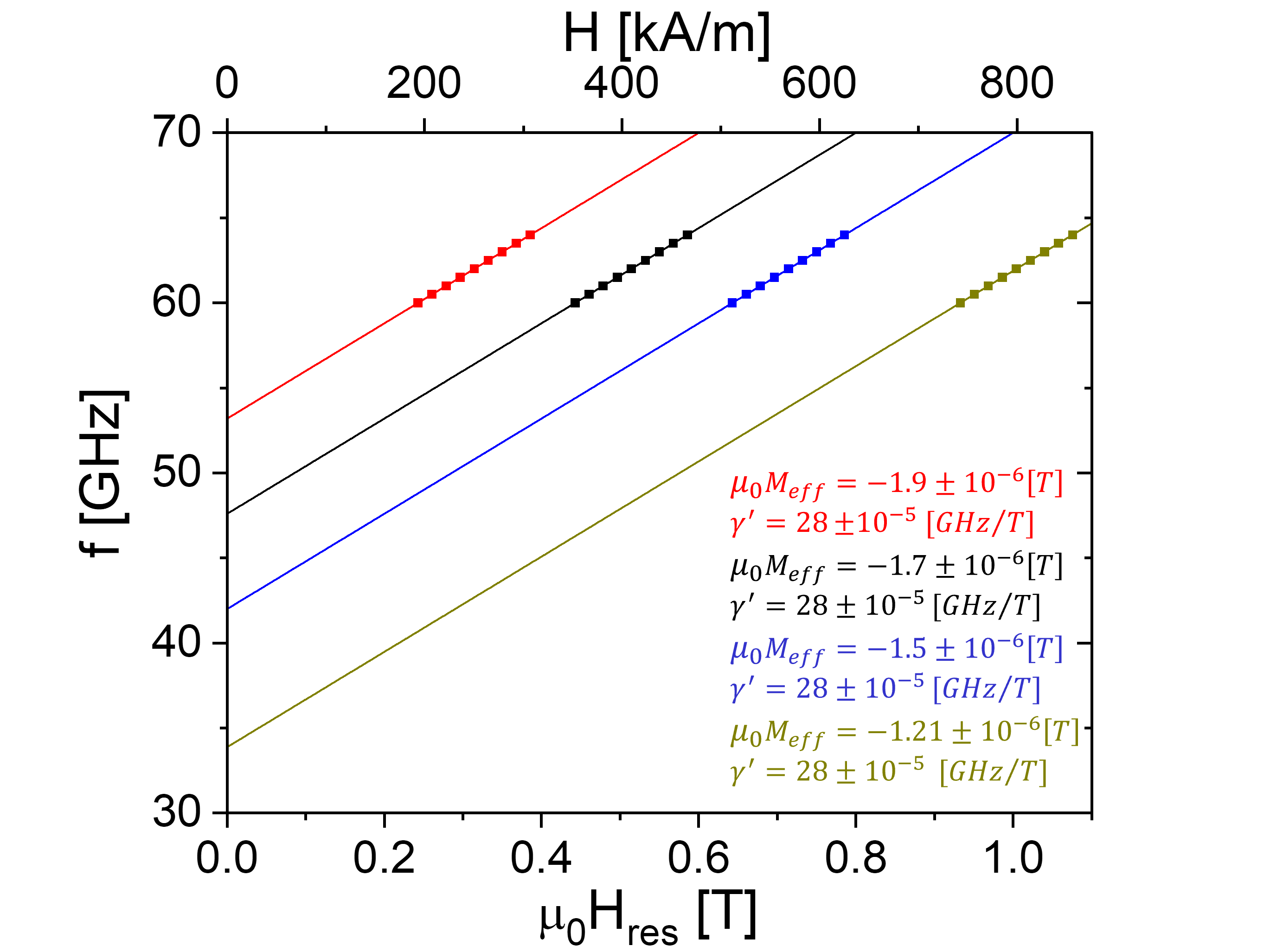}
     \end{subfigure}
    \begin{subfigure}[b]{0.4\textwidth}
         \centering
         \caption{$N=5$\hfill\,}
         \includegraphics[width=\textwidth,trim={0.9cm 0 2.5cm 0},clip]{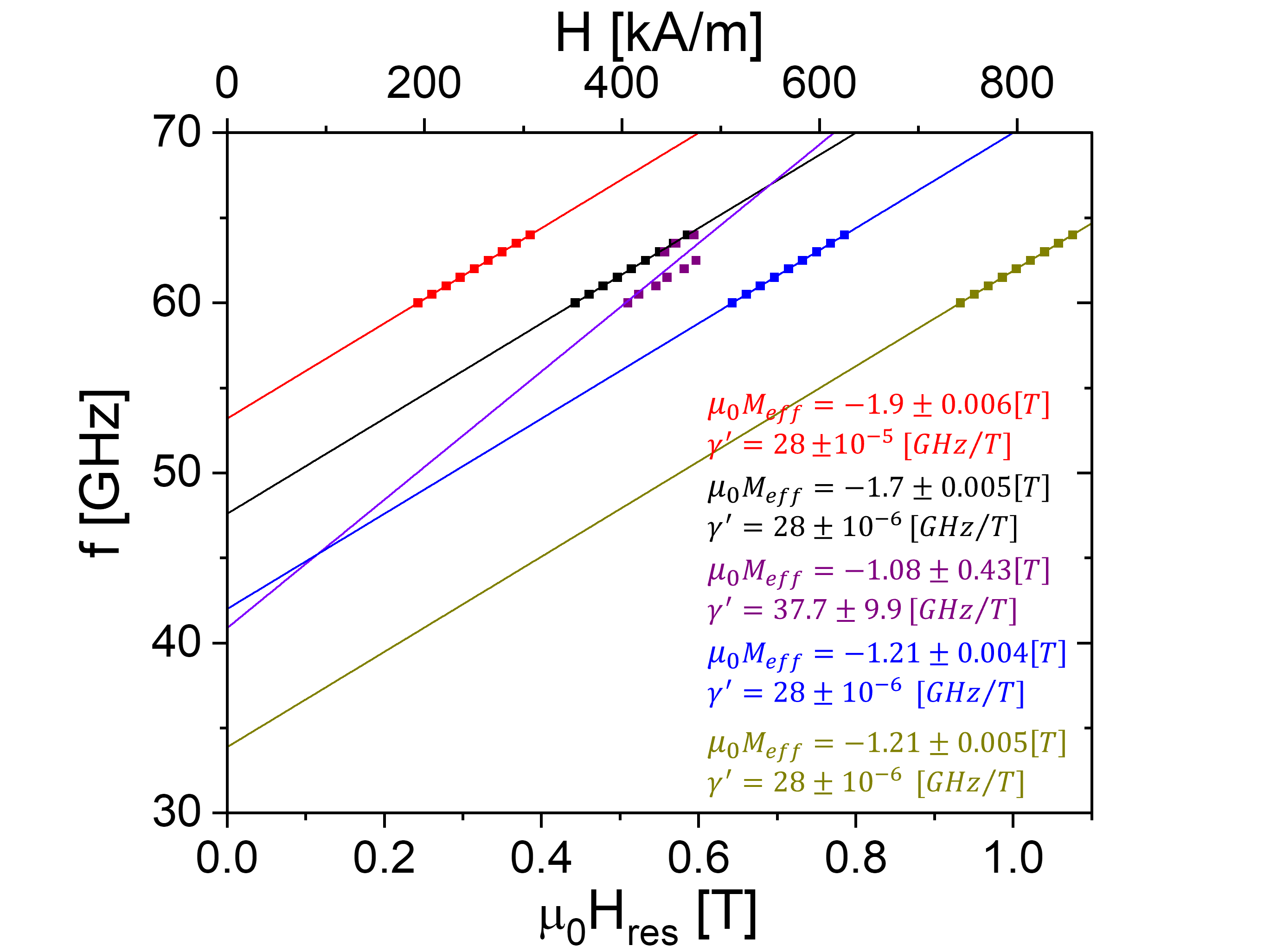}
     \end{subfigure}
    \hfill
        \caption{(a)-(e) Kittel plots based on the resonance field extracted from the simulated data using an increasing number of resonances $N=\{1,...,5\}$. The data in each plot was fitted using equation \eqref{eq:Kittel} and the results are indicated in each figure.}
        \label{fig:Kittel_vary_Meff}
\end{figure}

\begin{figure}
     \centering
     \begin{subfigure}[b]{0.48\textwidth}
         \caption{\hfill\,}
         \centering
         \includegraphics[width=\textwidth,trim={0.8cm 0 0.5cm 1.2cm},clip]{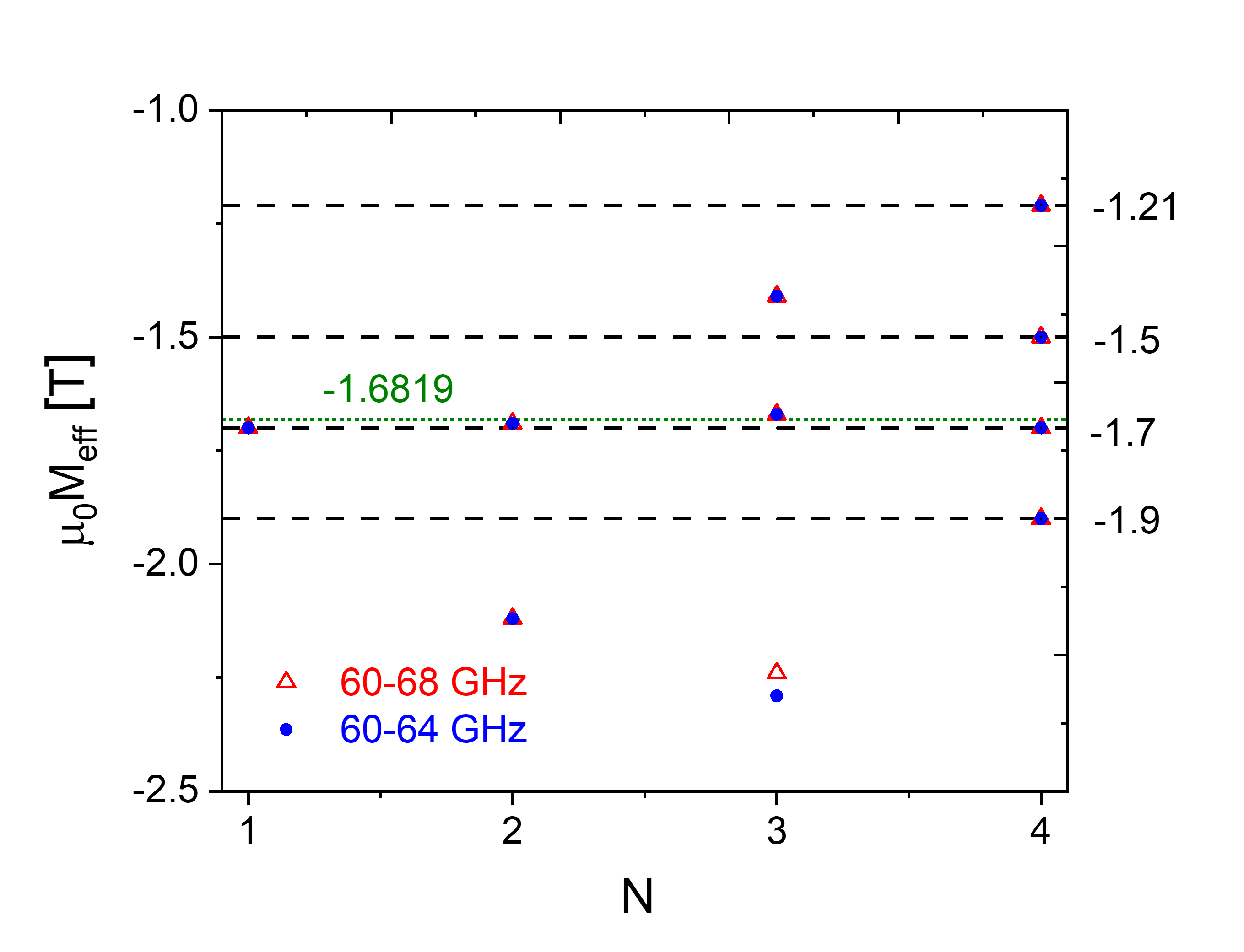}
     \end{subfigure}
     \hfill
     \begin{subfigure}[b]{0.48\textwidth}
         \centering
         \caption{\hfill\,}
         \includegraphics[width=\textwidth,trim={0.8cm 0 0.5cm 1.2cm},clip]{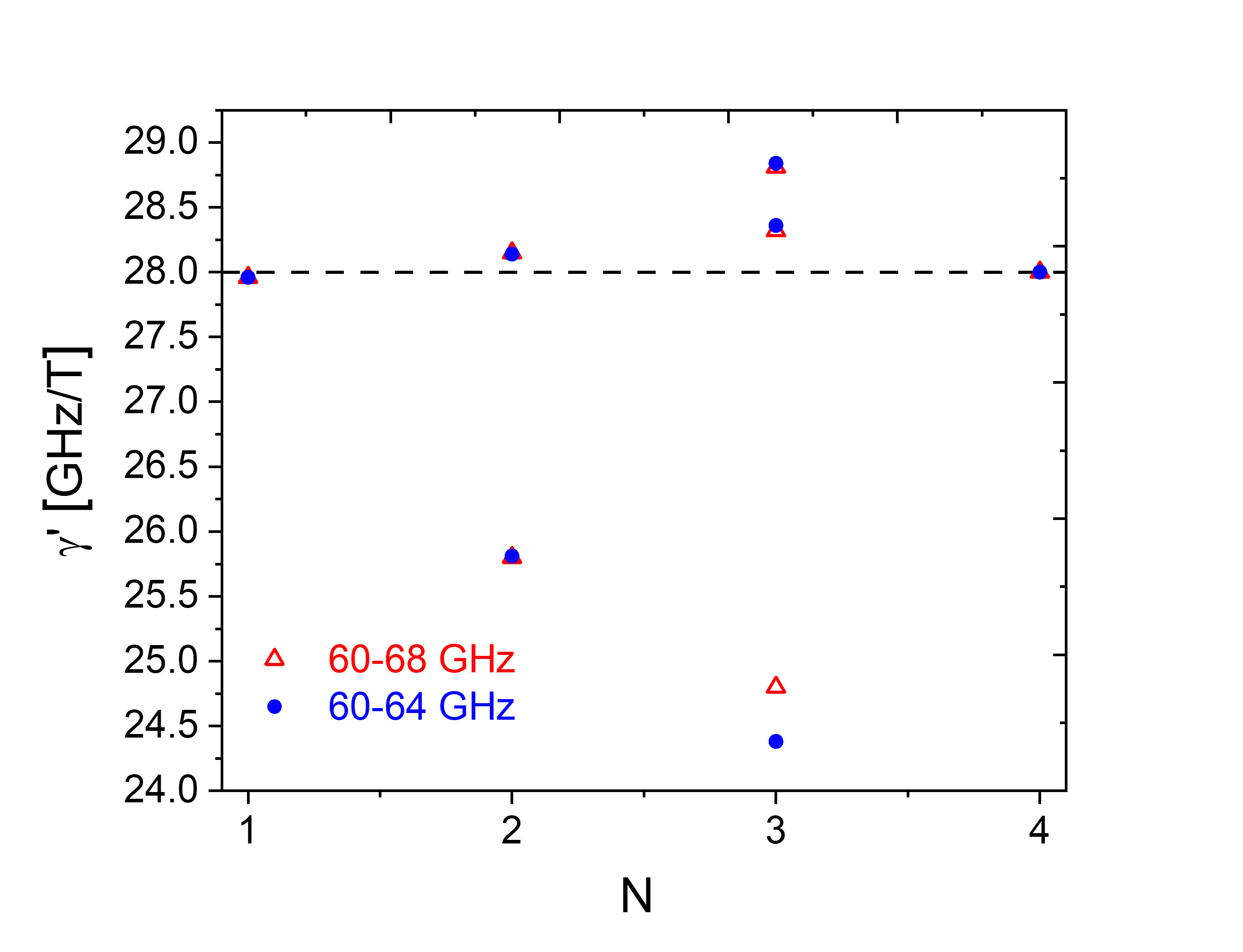}
     \end{subfigure}
        \caption{(a) Effective magnetization $M\iu{Fit}\is{eff,k}$ and (b) gyromagnetic ratio ${\gamma\iu{Fit}_{k}}'$ for $k=\{1,...,N\}$ as a function of the number of resonances $N=\{1,...,4\}$ used to fit the simulated data using equation \eqref{eq:Kittel}. The error-margins for all data points are smaller than the symbol size and are thus omitted. Blue symbols use spectra covering a frequency range from $60-64\, [GHz]$ (see figure \ref{fig:Kittel_vary_Meff}) whereas red symbols use an extended frequency range covering $60-68\, [GHz]$ (see text for details). The black dashed lines in both graphs represent the values of the four resonances used to simulate the data. The green dashed line represents the weighted average of the effective magnetization.}
        \label{fig:Summary_vary_Meff}
\end{figure}

For the simulations in this section we assumed that all four constituents of the material shared the same gyromagnetic ratio $\gamma_{k}'=28\, [\frac{GHz}{T}]$, with $k=\{1,...,4\}$. The shift between the individual resonance fields is therefore only caused by the difference in the effective magnetization of the different constituents. We have chosen $\mu_0M\is{eff,k}=\{-1.9,-1.7,-1.5,-1.21\}\, [T]$. In addition, we also assumed that all constituents share the same damping parameter $\alpha_k=0.01$. However, the inhomogeneous broadening was assumed to be different: $\mu_0\Delta H_{0,k}=\{200, 100, 75, 150\} [mT]$. The simulated resonances only contain an absorptive part and the amplitudes were $A_k=\{0.3, 1.0, 0.2, 0.1\}$. For the initial analysis we assumed that the frequency range for the microwave frequency $f$ ranged from $60\, [GHz]$ to $64\, [GHz]$ with spectra recorded in $0.5\, [GHz]$ intervals. The results of the simulations are shown in figure \ref{fig:multipleResonances}.\\
As described above, all spectra are fitted using different numbers of resonances $N$ in the fitting function. The spectra and results of the fits are shown in figure \ref{fig:ResonanceFits} exemplary for spectra with a microwave frequency $f=62\,[GHz]$. For fits that use less than the four resonances that are present in the spectra it is clear from the residuals (shown as insets) that the fit does not fully describe the data. However, noise present in experimental data will often make this less obvious, for more details on this see the later discussion regarding the influence of noise in section \ref{Noise}. As can be expected, when over fitting the data using $N=5$ resonances, the residual does only change slightly compared to $N=4$. The oscillations of the residuals for $N=4$ and $N=5$ are thus a result of numerical errors during the fitting process and the condition used to terminate the fitting algorithm.
\\
As shown in figure \ref{fig:adjustedR2} for all fits the adjusted-$R^2$ values are very close to 1. We therefore opted to plot the deviation of the adjusted-$R^2$ from 1 in this figure to make the deviations more easily accessible. As can be expected when over fitting the data by using $N=5$ resonances the adjusted-$R^2$ does not change significantly and hence we omit this data in the figure. 
\\
The resonance fields $H\iu{Fit}_{res,k}$ and their error margins $\sigma_{H\iu{Fit}_{res,k}}$ extracted from the fits of the raw data can now be used to determine the gyromagnetic ratio ${\gamma\iu{Fit}_{k}}'$ and the effective magnetization $M\iu{Fit}\is{eff,k}$ as well as their error margins, using equation \eqref{eq:Kittel}. The results are shown in figure \ref{fig:Kittel_vary_Meff} for $N=\{1,...,5\}$. In all cases except for $N=5$ the fits using equation \eqref{eq:Kittel} describe the relationship between microwave frequency and extracted resonance fields very well. Even close inspection of the fits provides no evidence that for $N=\{1,..,3\}$ the data is missing one or more resonances present in the material. For the overfitted data using $N=5$ the four resonances that are present in the material are captured accurately. However, the additional fictional resonance features a sudden shift and is therefore not well described by equation \eqref{eq:Kittel}. This together with for example the lack of an improvement of the adjusted-$R^2$ compared to the $N=4$ fit should provide clear indications that the additional resonance of this fit is an artifact of overfitting. We will therefore exclude this data from subsequent discussions.
\\
The results of the fits shown in the Kittel plots in figure \ref{fig:Kittel_vary_Meff} are summarized in figure \ref{fig:Summary_vary_Meff} as blue symbols. This figure also contains results using an extended frequency range where we used the same methodology but simulated spectra over a microwave frequency range from $60\, [GHz]$ to $68\, [GHz]$ with spectra recorded in $1\, [GHz]$ intervals. The statistical error-margins in both graphs of this figure are smaller than the size of the symbols. For comparison the values for the effective magnetization and gyromagnetic ratio used to simulate the spectra are shown as black dashed lines. \\
One of the key observations is that when choosing fewer resonances to fit the data than are present in the spectra, the fitted values for both the effective magnetization and the gyromagnetic ratio deviate significantly from their true values. However, the corresponding Kittel plots and the adjusted-$R^2$ values give very little indication that these values may not be trustworthy. If we consider for example the case where the fit used $N=3$ resonances the fit lines in the corresponding Kittel plot (see figure \ref{fig:Kittel_vary_Meff} (c)) describe the data very well and the adjusted-$R^2$ values of the fitted spectra are extremely close to 1 (see figure \ref{fig:adjustedR2}). Because we have not added noise to the simulated spectra one can detect the presence of the additional resonance by examining the simulated spectrum and the corresponding fit (see figure \ref{fig:ResonanceFits} (c)) or the residual (inset of the same figure). Furthermore, for the $N=3$ case the fitted values for the effective magnetization do not agree with any of the true values of the constituent materials (see figure \ref{fig:Summary_vary_Meff}). One of the $N=3$ effective magnetization values is entirely out of the range of values used in the simulation. It is noteworthy that in this case the corresponding gyromagnetic ratio also significantly underestimates the true gyromagnetic ratio of the material.\\ 
The limiting case where the spectra were fitted with a single resonance $(N=1)$ leads to a fitted gyromagnetic ratio of  ${\gamma\iu{Fit}}'=27.96\pm 0.01 [\frac{GHz}{T}]$ that at least comes close to the true value $\gamma_{k}'=28\, [\frac{GHz}{T}]$ used for all resonances. However, the error-margins obtained from the fit are still smaller than the observed deviation. For the effective magnetization in this case one can compare the fitted value $\mu_0M\iu{Fit}\is{eff}=-1.692\pm 0.001\,[T]$ with the weighted mean of the effective magnetization $\overline{M}\is{eff}$ of the four resonances present in the spectra:
\begin{equation}
 \overline{M}\is{eff}=\frac{\sum\limits_{k=1}^4A_k M\is{eff,k}}{\sum\limits_{k=1}^4A_k}. 
\end{equation}
For the simulated spectra one has $\mu_0\overline{M}\is{eff}=-1.6819\,[T]$ which is shown as a dashed green line in figure \ref{fig:Summary_vary_Meff} (b). One observes that the fitted value is close to the weighted mean but its error-margins are again smaller than the deviation. 
\\
We conclude this section by noting that doubling the frequency range of the simulations causes small differences in the extracted values obtained but it does not fundamentally change them (see figure \ref{fig:Summary_vary_Meff}). 
\subsection{Constituents with different gyromagnetic ratio}\label{gamma-variation}
\begin{figure}
     \centering
     \begin{subfigure}[b]{0.48\textwidth}
         \caption{\hfill\,}
         \centering
         \includegraphics[width=\textwidth,trim={0.1cm 0cm 0.5cm 1cm},clip]{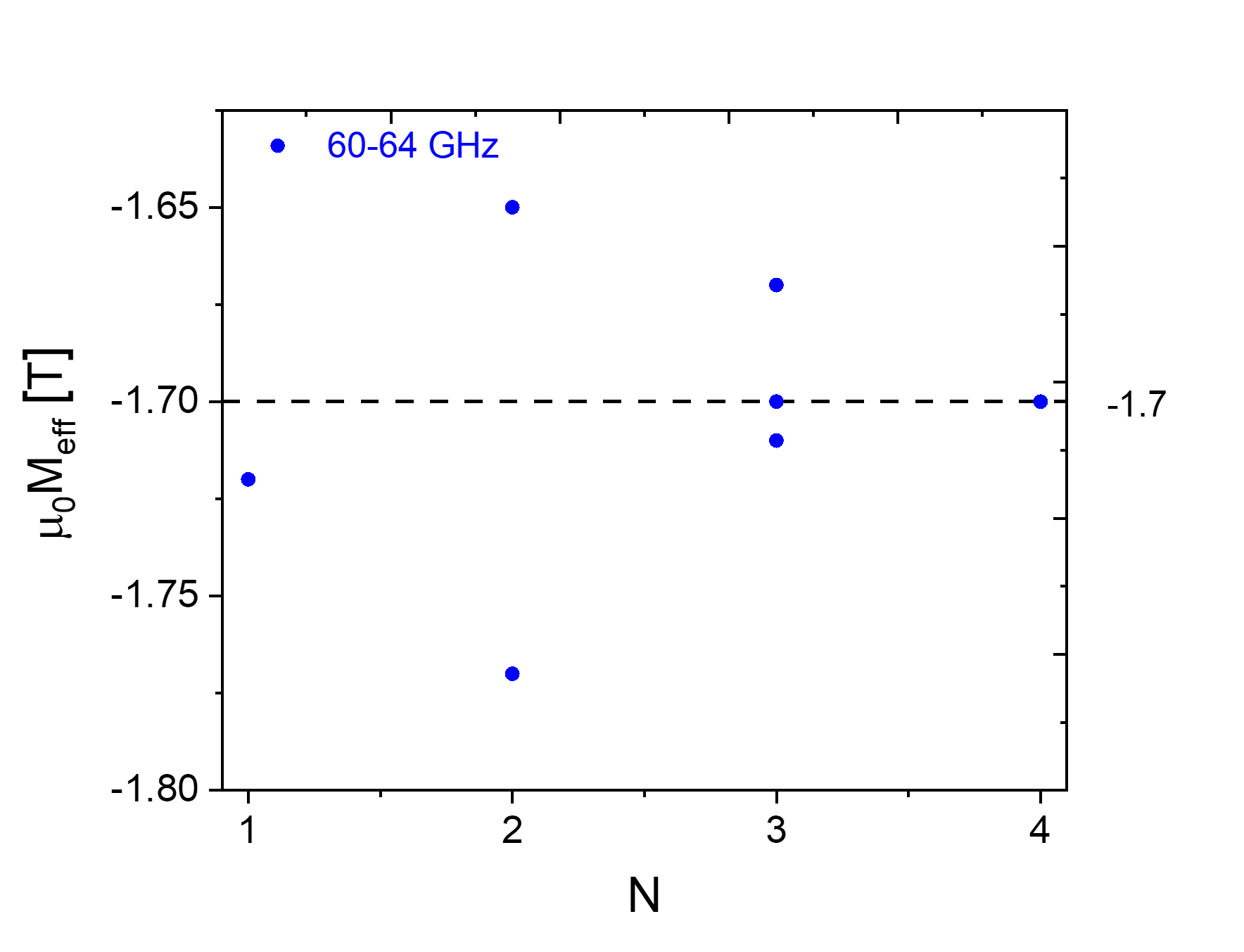}
     \end{subfigure}
     \hfill
     \begin{subfigure}[b]{0.48\textwidth}
         \centering
         \caption{\hfill\,}
         \includegraphics[width=\textwidth,trim={0cm 0cm 0.5cm 1cm},clip]{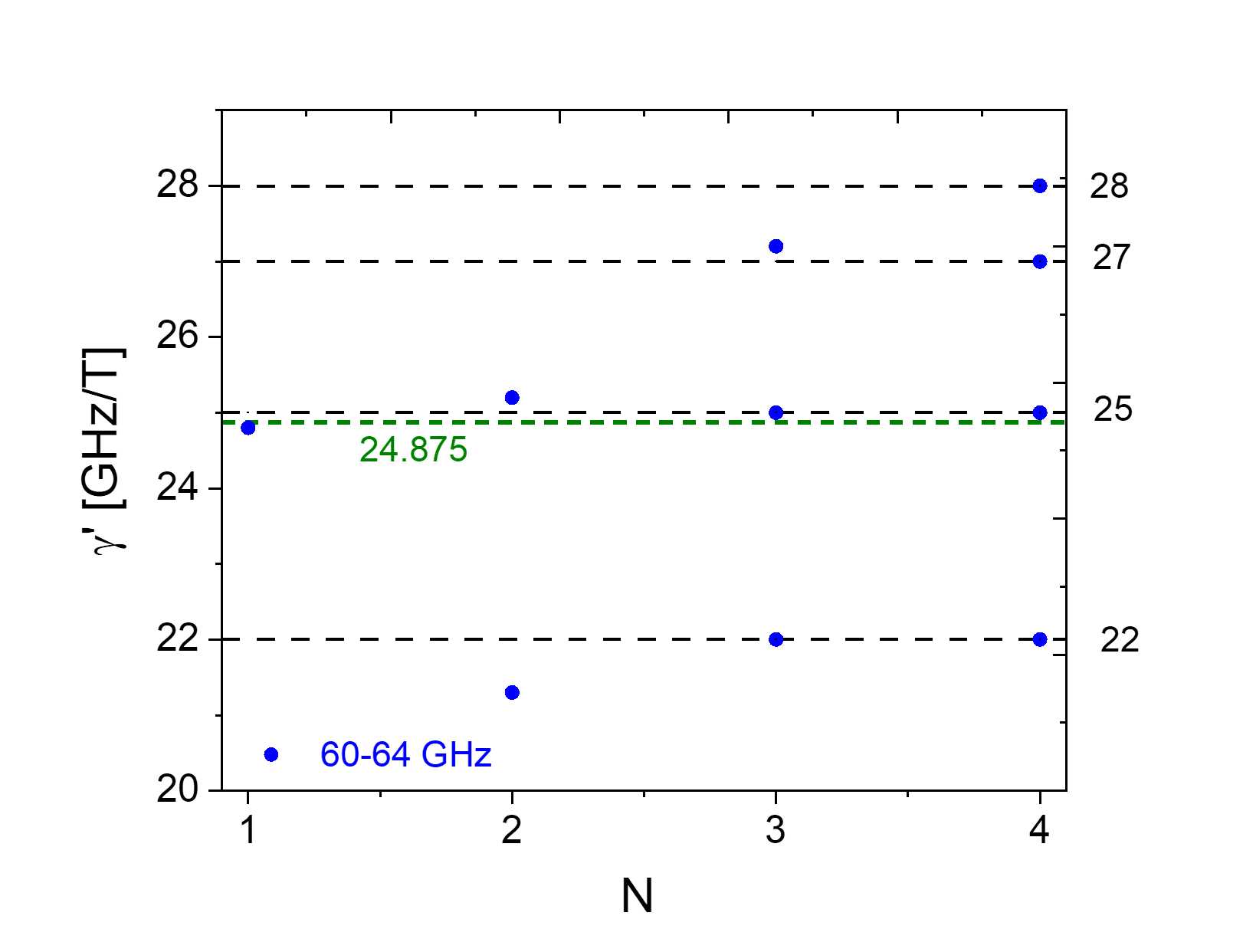}
     \end{subfigure}
        \caption{(a) Effective magnetization $M\iu{Fit}\is{eff,k}$ and (b) gyromagnetic ratio ${\gamma\iu{Fit}_{k}}'$ for $k=\{1,...,N\}$ as a function of the number of resonances $N=\{1,...,5\}$ used to fit the simulated data using equation \eqref{eq:Kittel}. The error-margins for all data points are smaller than the symbol size and are thus omitted. Blue symbols use spectra covering a frequency range from $60-64\, [GHz]$. The black dashed lines in both graphs represent the values of the four resonances used to simulate the data. The green dashed line represents the weighted average of the gyromagnetic ratio.}
        \label{fig:Summary_vary_gamma}
\end{figure}

For the simulations in this section we assumed that all four constituents of the material shared the same effective magnetization $\mu_0M\is{eff,k}=-1.7\, [T]$, with $k=\{1,...,4\}$. The shift between the individual resonance fields is therefore now only caused by the differences in the gyromagnetic ratio of the different constituents. We have chosen $\gamma_{k}'=\{22,25,27,28\}\, [\frac{GHz}{T}]$. We assumed again that all constituents share the same damping parameter $\alpha_k=0.01$ and for the inhomogeneous broadening we used $\mu_0\Delta H_{0,k}=\{200, 100, 75, 150\} [mT]$. As before the simulated resonances only contain an absorptive part and the amplitudes were $A_k=\{0.3, 1.0, 0.2, 0.1\}$. For the analysis we assumed that the frequency range for the microwave frequency $f$ ranged from $60\, [GHz]$ to $64\, [GHz]$ with spectra recorded in $0.5\, [GHz]$ intervals.\\
The data analysis follows the same methodology as described in the previous section. The final results are summarized in figure \ref{fig:Summary_vary_gamma}.
\\
 The statistical error-margins in both graphs of this figure are again smaller than the size of the symbols and for comparison the values for the effective magnetization and gyromagnetic ratio used to simulate the spectra are shown as black dashed lines. The observations are very similar to those discussed in section \ref{Meff-Variation}. When choosing more than one but fewer resonances than are present in the spectra to fit the data, the fitted values for both the effective magnetization and the gyromagnetic ratio deviate significantly from their true values. The results for a single resonance fit are again close to the true value of the effective magnetization and the weighted mean of the gyromagnetic ratio. As before, they deviate slightly more than their statistical error-margins from the true value of the effective magnetization and the weighted mean of the gyromagnetic ratio. While this deviation is particularly obvious for the effective magnetization as shown in figure \ref{fig:Summary_vary_gamma} (a), we note that the y-axis in this plot covers a rather small range of values and the relative deviation from the true value remains small. 
\subsection{Influence of noise}\label{Noise}
\begin{figure}
     \centering
     \begin{subfigure}[b]{0.48\textwidth}
         \caption{$N=3$ fit, $\sigma_N=2\cdot10^{-3}$\hfill\,}
         \centering
         \includegraphics[width=\textwidth,trim={0 0 0 0.65cm},clip]{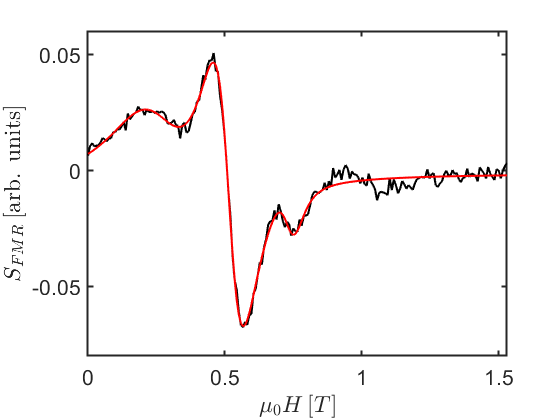}
     \end{subfigure}
     \begin{subfigure}[b]{0.48\textwidth}
         \centering
         \caption{$N=3$ residual, $\sigma_N=2\cdot10^{-3}$\hfill\,}
         \includegraphics[width=\textwidth,trim={0 0 0 0.65cm},clip]{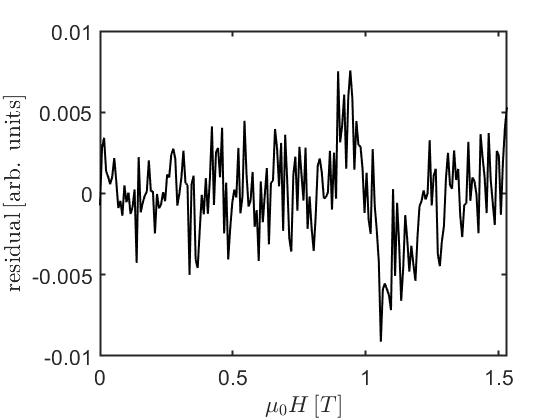}
     \end{subfigure}
        \caption{Exemplary spectrum (black) and fits (red) for $f=62\,[GHz]$ (a) and the corresponding residual (b). The simulation uses a standard deviation $\sigma_N=2\cdot10^{-3}$ for the normal distribution of the noise added to the signal.}
        \label{fig:Noise_fits}
\end{figure}

\begin{figure}
     \centering
     \begin{subfigure}[b]{0.48\textwidth}
         \caption{\hfill\,}
         \centering
         \includegraphics[width=\textwidth,trim={0.6cm 0 2cm 1.2cm},clip]{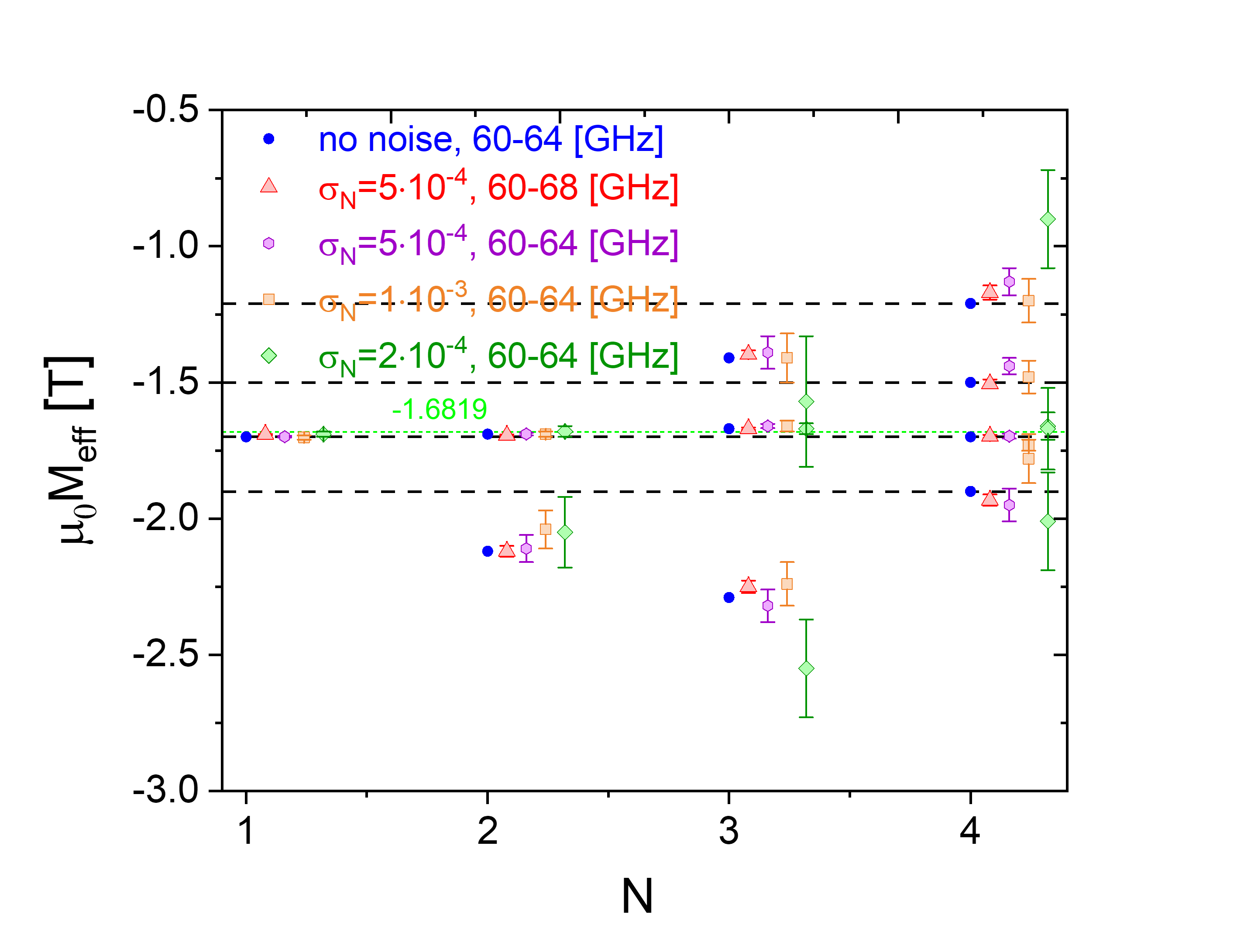}
     \end{subfigure}
     \hfill
     \begin{subfigure}[b]{0.48\textwidth}
         \centering
         \caption{\hfill\,}
         \includegraphics[width=\textwidth,trim={0.6cm 0 2cm 1.2cm},clip]{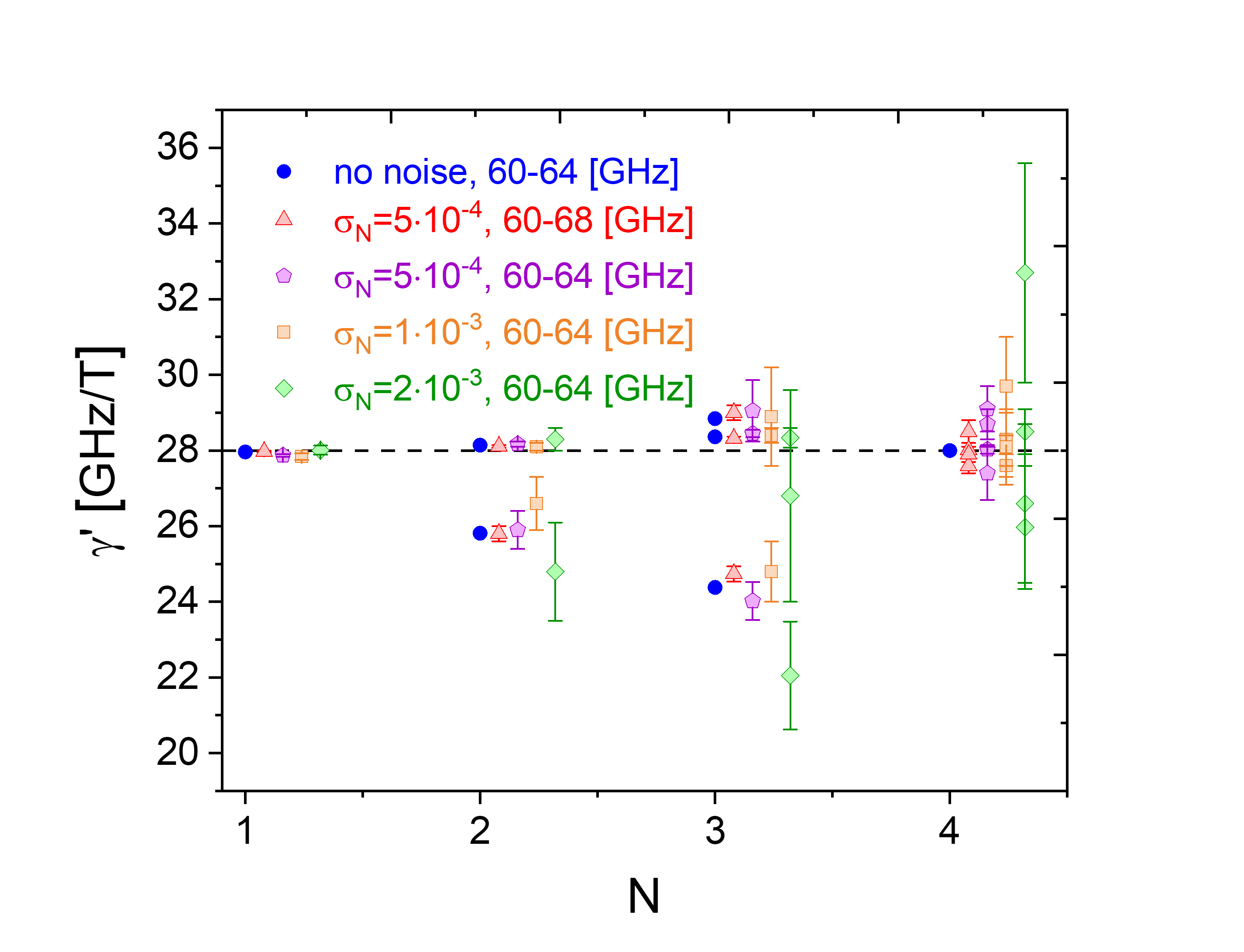}
     \end{subfigure}
        \caption{(a) Effective magnetization $M\iu{Fit}\is{eff,k}$ and (b) gyromagnetic ratio ${\gamma\iu{Fit}_{k}}'$ for $k=\{1,...,N\}$ as a function of the number of resonances $N=\{1,...,4\}$ used to fit the simulated data using equation \eqref{eq:Kittel}. The black dashed lines in both graphs represent the values of the four resonances used to simulate the data. The green dashed line represents the weighted average of the effective magnetization. The standard deviations $\sigma_N$ used for the normal distributed noise added to the spectra is indicated in the legend. To improve readability of the graphs the data sets are slightly shifted to the right with increasing noise level. All but one data set shown here use a frequency range for the spectra from $60\, [GHz]$ to $64\, [GHz]$. The exception is the data shown as red triangles, which uses a broader frequency range covering $60\, [GHz]$ to $68\, [GHz]$. }
        \label{fig:Summary_vary_Meff_Noise}
\end{figure}

\begin{figure}
     \centering
     \begin{subfigure}[b]{0.48\textwidth}
         \caption{\hfill\,}
         \centering
         \includegraphics[width=\textwidth,trim={0.6cm 0 0.2cm 1.2cm},clip]{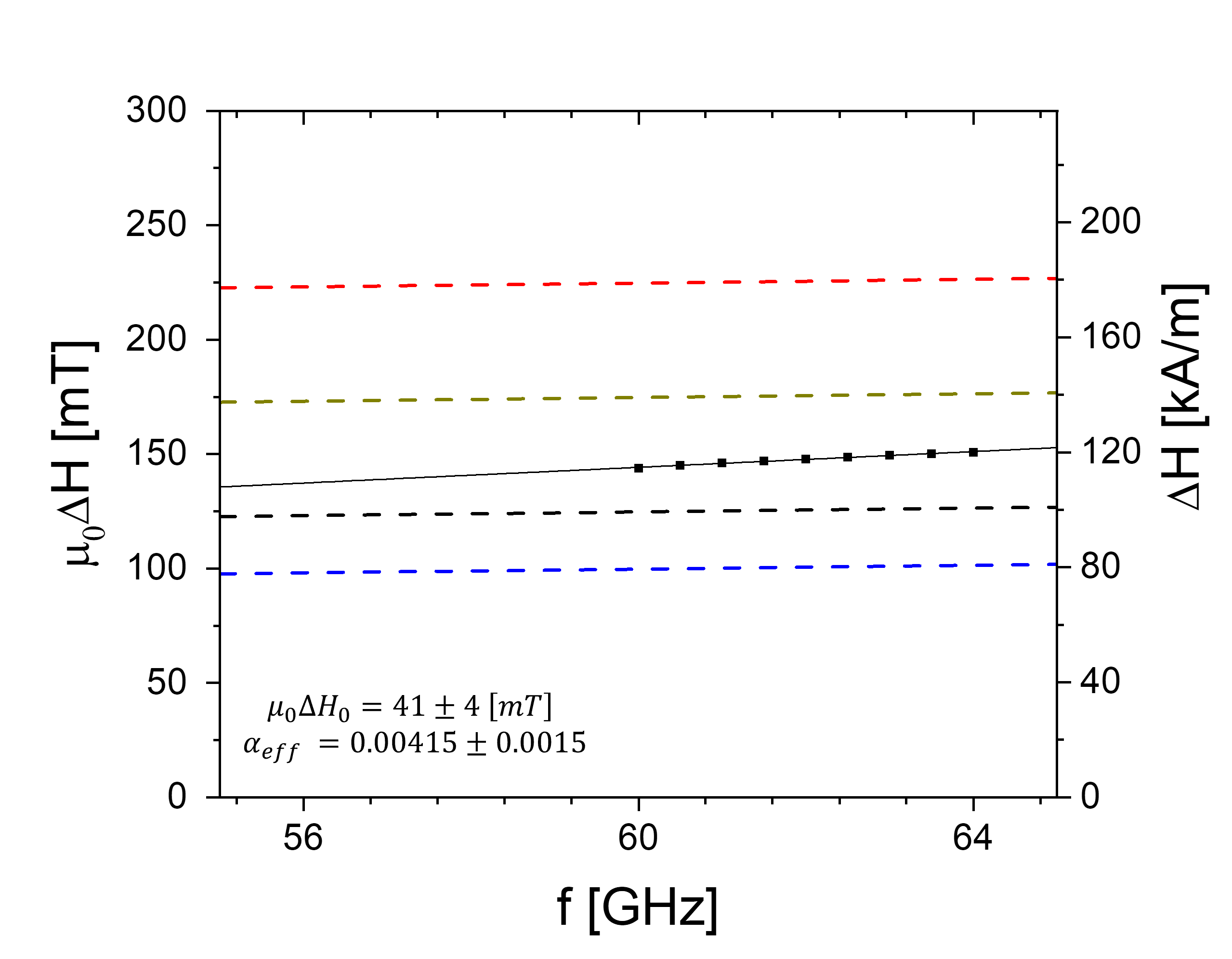}
     \end{subfigure}
     \hfill
     \begin{subfigure}[b]{0.48\textwidth}
         \centering
         \caption{\hfill\,}
         \includegraphics[width=\textwidth,trim={0.6cm 0 0.2cm 1.2cm},clip]{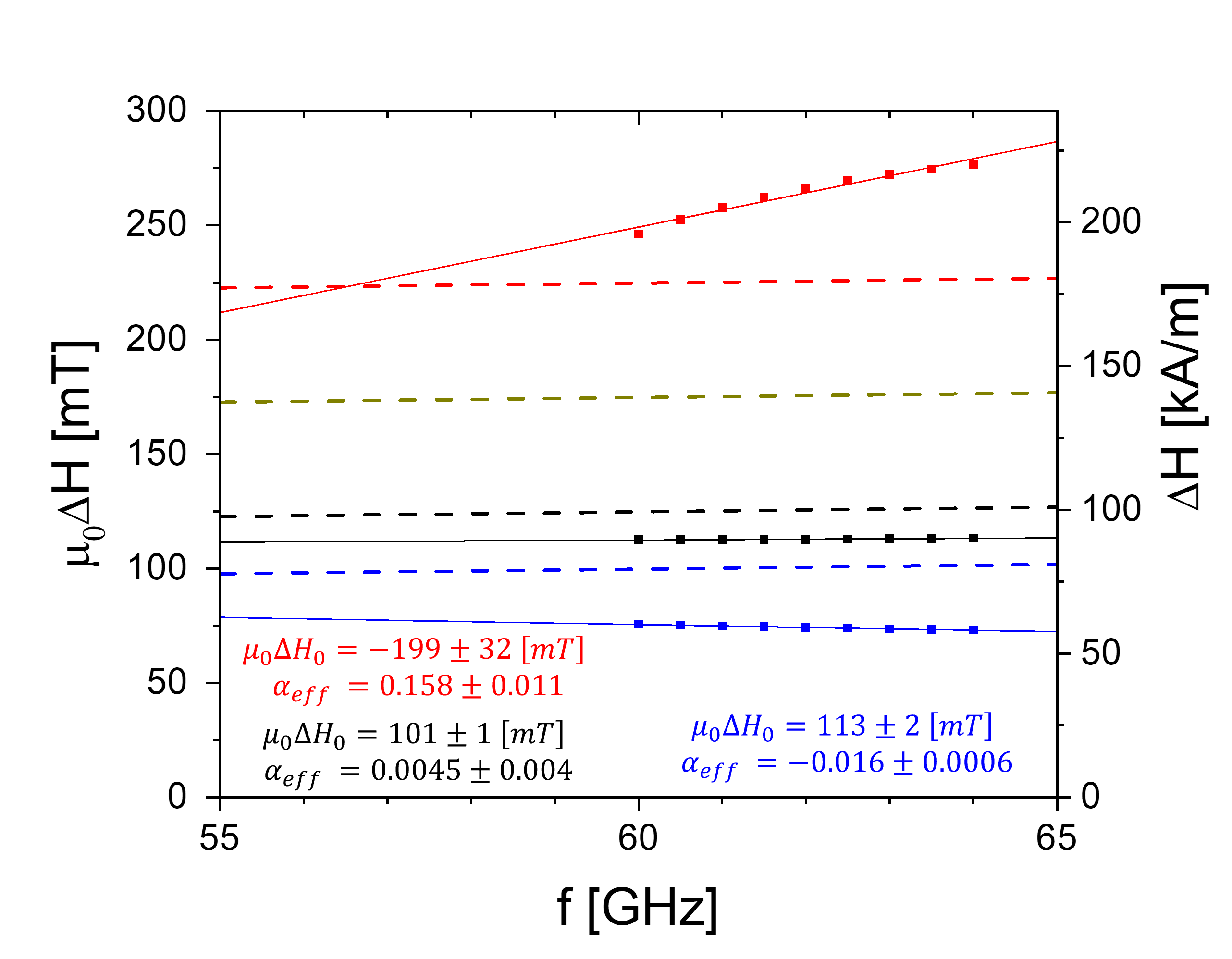}
     \end{subfigure}
    \begin{subfigure}[b]{0.48\textwidth}
         \centering
         \caption{\hfill\,}
         \includegraphics[width=\textwidth,trim={0.6cm 0 0.2cm 1.2cm},clip]{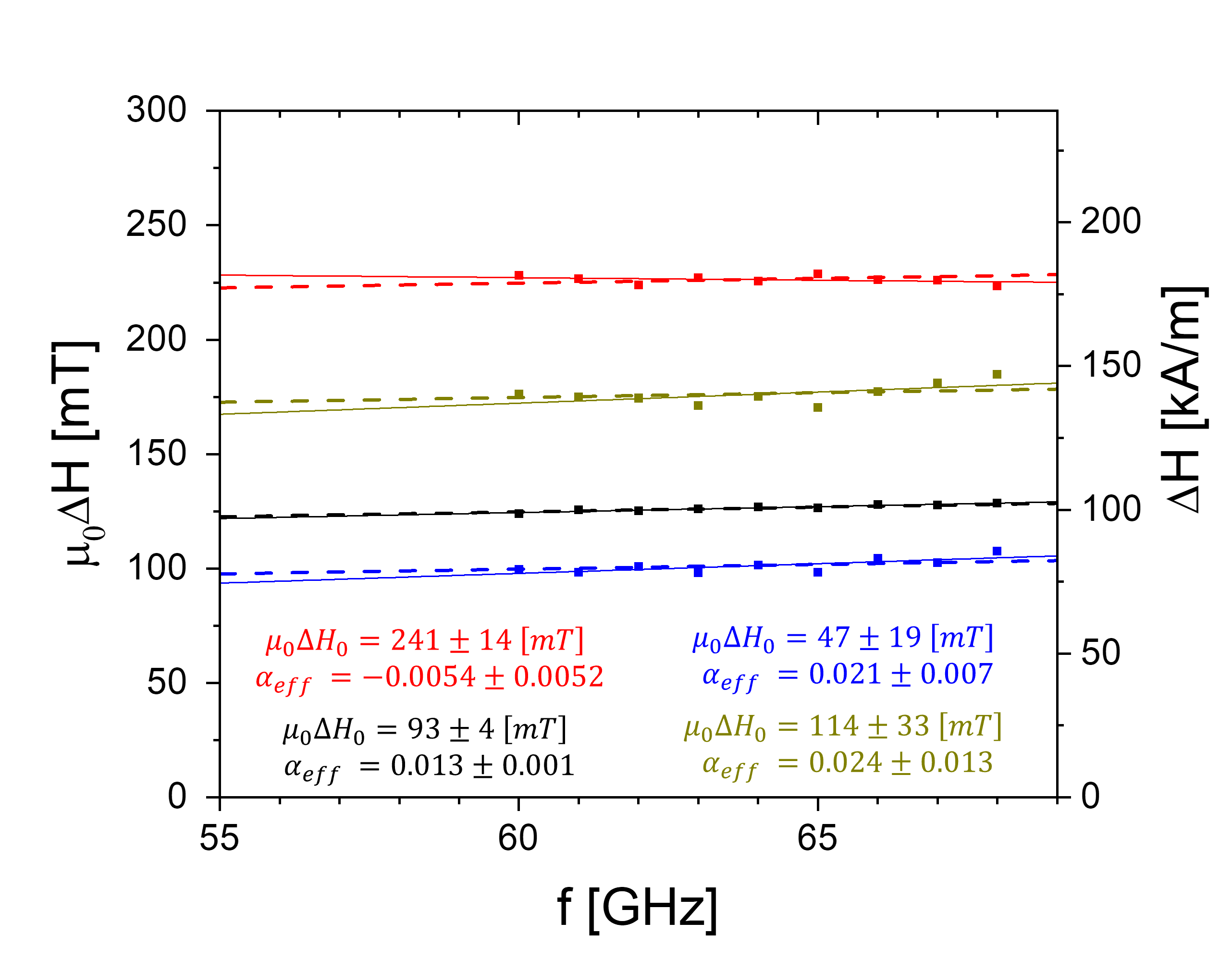}
     \end{subfigure}
        \caption{Linewidth as a function of microwave frequency extracted from spectra using (a) a single resonance $N=1$, (b) three resonances $N=3$, and (c) four resonances $N=4$. The simulated data in (a) \& (b) had no noise added and covered a frequency range from $60\, [GHz]$ to $64\, [GHz]$, the data in (c) had noise with a standard deviations $\sigma_N=5\cdot10^{-4}$ added and covered a frequency range from $60\, [GHz]$ to $68\, [GHz]$. The dashed lines in all graphs represent the frequency dependence of the linewidth used for the simulations.}
        \label{fig:linewidthPlots}
\end{figure}

\begin{figure}
     \centering
     \begin{subfigure}[b]{0.48\textwidth}
         \caption{\hfill\,}
         \centering
         \includegraphics[width=\textwidth,trim={0.6cm 0 1.2cm 1.2cm},clip]{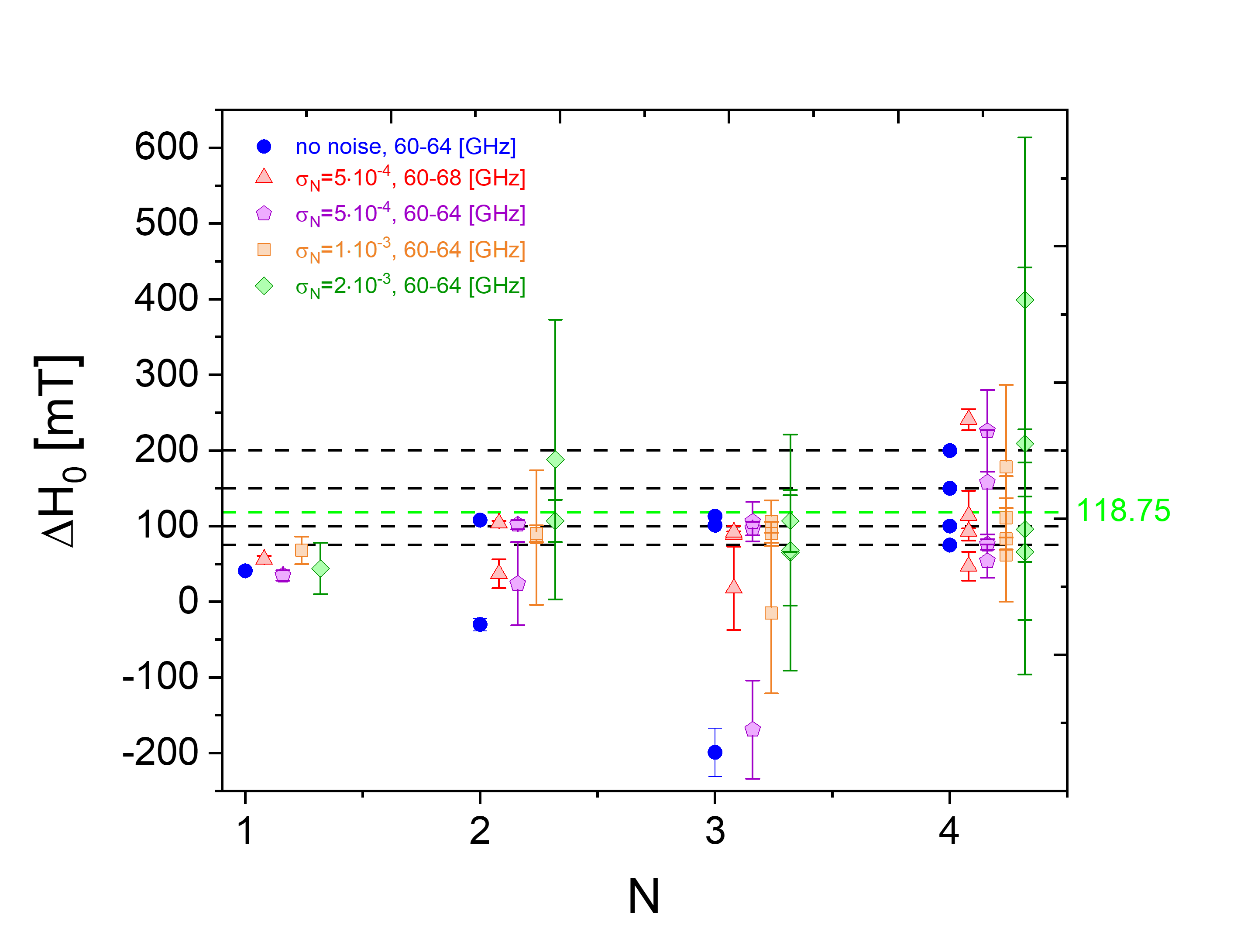}
     \end{subfigure}
     \hfill
     \begin{subfigure}[b]{0.48\textwidth}
         \centering
         \caption{\hfill\,}
         \includegraphics[width=\textwidth,trim={0.6cm 0 1.2cm 1.2cm},clip]{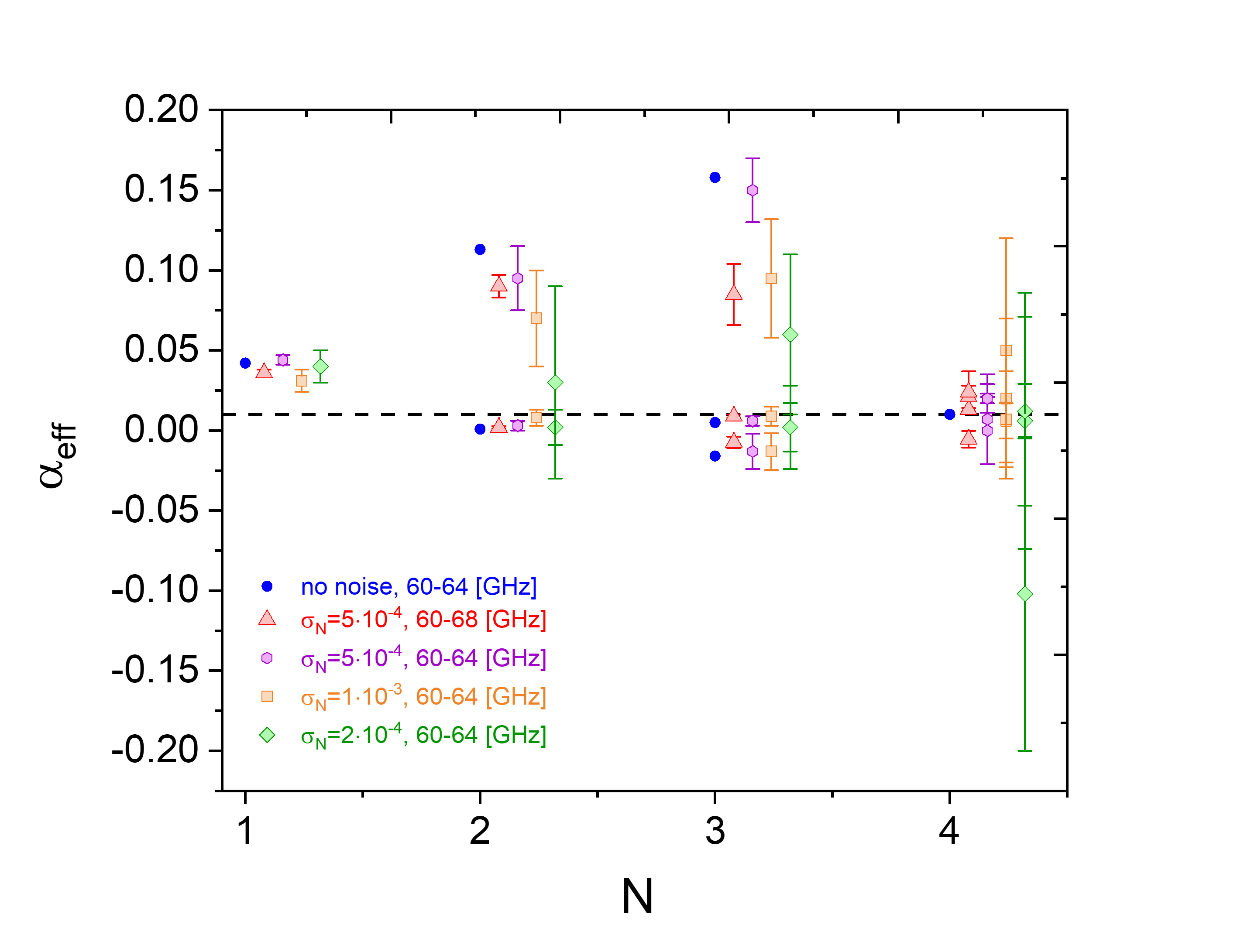}
     \end{subfigure}
        \caption{(a) Inhomogeneous linewidth broadening $\Delta H_{0,k}$   and (b) damping parameter $\alpha_k$ for $k=\{1,...,N\}$ as a function of the number of resonances $N=\{1,...,4\}$ used to fit the simulated linewidth data using equation \eqref{eq:linewidth}. The black dashed lines in both graphs represent the values of the four resonances used to simulate the data. The green dashed line represents the weighted average of the inhomogeneous linewidth broadening. The standard deviations $\sigma_N$ used for the normal distributed noise added to the spectra is indicated in the legend. To improve readability of the graphs the data sets are slightly shifted to the right with increasing noise level. All but one data set shown here use a frequency range for the spectra from $60\, [GHz]$ to $64\, [GHz]$. The exception is the data shown as red triangles, which uses a broader frequency range covering $60\, [GHz]$ to $68\, [GHz]$. }
        \label{fig:dampingSummary}
\end{figure}

To investigate the influence of noise on the results for parameter values obtained after data analysis, we assumed in this section that the constituents of the material differed regarding their effective magnetizations but shared the same gyromagnetic ratio. We used the same parameters for the simulations as we have used in this case earlier, see section \ref{Meff-Variation}. To simulate noise, we added normally distributed random noise with zero mean, i.e. $\mu=0$, and different standard deviations $\sigma_N=\{5\cdot10^{-4},1\cdot10^{-3},2\cdot10^{-3}\}$ to the signal. The data analysis was done using exactly the same methodology as described before. One important difference regarding the fits of the spectra compared to the case shown in figure \ref{fig:ResonanceFits} is that when fitting the data with $N=4$ the residuals show no systematic variation with field but solely random noise. However, for fits with $N$ less than the actual number of resonances the residuals still reveal the presence of an additional resonance. This is shown exemplary in figure \ref{fig:Noise_fits} for a fit using $N=3$ resonances. 
\\
With increasing noise level of the spectra the fitting algorithm has increasing difficulty to properly fit all resonances. The error-margins of the fit parameters extracted from the spectra therefore also increase. These errors then propagate to the effective magnetization $M\iu{Fit}\is{eff,k}$ and gyromagnetic ratio ${\gamma\iu{Fit}_{k}}'$ derived by fitting equation \ref{eq:Kittel} to the broadband data. Figure \ref{fig:Summary_vary_Meff_Noise} summarizes the influence of noise on these parameters.
\\
As can be expected with increasing noise, the error-margins of the extracted values for the effective magnetization and the gyromagnetic ratio increase. For fits using a single resonance $N=1$ the results agree well with the results obtained without any noise and are close to the weighted mean of the effective magnetization and the true value of the gyromagnetic ratio. Because the amplitude $A_k$ of the strongest resonance is more than three times larger than the one with the next largest amplitude adding noise to the data does not fundamentally change the behavior of the single resonance fit of the individual spectra. Hence one can expect the single resonance fit to be relatively insensitive to noise as long as the strongest resonance remains clearly observable. \\
Consequently it is not surprising that for $N>1$ the resonances with smaller amplitudes either are missed entirely when fitting the individual spectra and/or lead to larger error margins in the values for the effective magnetization and the gyromagnetic ratio. For the effective magnetization and gyromagnetic ratio the behavior for $N=2$ and $N=3$ mimics that observed for spectra without noise. The parameter values are comparable to those obtained without noise but are more scattered due to the noise in the spectra. \\
The behavior of the fitted values for the gyromagnetic ratio for $N=4$ is interesting. In the case of noiseless data the fit exactly reproduces the gyromagnetic ratio used in the simulation. However, for noisy data one observes a relatively large spread. This is noticeable even for the lowest levels of noise and is accompanied with corresponding large error-margins, see figure \ref{fig:Summary_vary_Meff_Noise} (b). The deviations from the true value and the uncertainty of the gyromagnetic ratio also lead to deviations and increased uncertainty of the effective magnetization, see figure \ref{fig:Summary_vary_Meff_Noise} (a). This is where an increased frequency range of broadband ferromagnetic resonance spectroscopy can make a difference. To illustrate this we have included a simulation with a standard deviation of the noise $\sigma_N=5\cdot10^{-4}$ and otherwise identical parameters, but using a frequency range for $f$ from $60\, [GHz]$ to $68\, [GHz]$ with spectra recorded in $1\, [GHz]$ intervals. Note that we are thus using the same number of spectra as in the cases covering a frequency range from $60\, [GHz]$ to $64\, [GHz]$, they are simply more spread out in frequency. This data is shown as red triangles in figure \ref{fig:Summary_vary_Meff_Noise}. The increased frequency range results in a significantly smaller spread of the gyromagnetic ratio and consequently also leads to values for the effective magnetization that are closer to the true values. This observation is consistent with earlier work for materials with a single resonance that used an asymptotic approach to further improve the accuracy for the extraction of the gyromagnetic ratio from broadband FMR measurements \cite{Shaw2013}.
\\
For this example we also investigated the influence noise and model assumptions on the extraction of the Gilbert damping parameter $\alpha_k$ and the inhomogeneous linewidth broadening $\Delta H_{0,k}$. In order to extract these parameter from the simulated data one can plot the linewidth $\Delta H_k$ as a function of the microwave frequency $f$ and fit this data using equation \eqref{eq:linewidth}. This is shown exemplary in figure \ref{fig:linewidthPlots} for a few cases.
\\
In part (a) of this figure we have used simulation data without noise and a single resonance fit $N=1$ to extract the linewidth. While the data is well described by equation \eqref{eq:linewidth} neither the extracted damping parameter nor the inhomogeneous broadening reflects the properties of any of the constituents or their weighted means. This becomes even more obvious when using three resonances to fit the noiseless data as shown in figure \ref{fig:linewidthPlots} (b). Here one would obtain an unphysical negative inhomogeneous linewidth contribution for one of the resonances. For this data the careful observer may realize that the data shows some curvature that is not captured by the linear fit. However, this can easily missed in particular if noise is present in the data. For another resonance in the same figure the damping parameter would be negative and thus unphysical. Figure \ref{fig:linewidthPlots} (c) shows a plot of the linewidth using the correct number of resonances $N=4$ for spectra with noise with a standard deviation $\sigma_N=5\cdot10^{-4}$. Despite using the simulation data covering the broader frequency range one can still encounter fits using equation \eqref{eq:linewidth} that might suggest a negative damping (red data). We would like to point out that given the relatively small damping parameter and the large inhomogeneous broadening we assumed for the simulations they represent somewhat of a worst case scenario and one would therefore need a much larger frequency range to accurately determine the damping parameters and the inhomogeneous linewidth broadening. However, this example clearly demonstrates that extra care has to be taken when attempting to extract damping related parameters using equation \eqref{eq:linewidth} from spectra that contain a known or unknown number of resonances. After all, whether the damping parameter is large or small is not known a priori. Minimizing the noise in the data and maximizing the frequency range are key to minimizing the error margins of the extracted parameters. For completeness we show in figure \ref{fig:dampingSummary} a summary of the results of the linewidth analysis.
\\
The most important result from the graphs in this figure is that despite using the correct number of resonances that produced reliable results for the effective magnetization and the gyromagnetic ratio the same data analysis can lead to significant deviations of the inhomogeneous linewidth and damping parameter of the fitted values from the true values of the material. Furthermore, if an incorrect number of resonances is used to fit the spectra wide variations of the parameters extracted from the frequency dependence of the linewidth can be expected. In this case the extracted parameters have no discernible relationship with the true values of the material even without noise present in the simulated data.  
\section{Conclusion}
In summary, we have highlighted some of the pitfalls that one can encounter when using broadband ferromagnetic resonance spectroscopy to characterize materials that have more than a single constituent and therefore exhibit multiple possibly overlapping resonances. \\
Our results show that it is desirable to have independent knowledge regarding the number of resonances that are expected in a material to avoid analyzing the data using an incorrect number of resonances. Our simulations have shown, even in the absence of noise, that assuming an incorrect number of resonances for the fit can result in the extraction of material parameters that are not consistent with any of the constituents present in the material. In these cases one observes deviations that far exceed the statistical error-margins associated with fitting the data. That is, the systematic error of choosing an incorrect number of resonances to describe the observed spectra by far exceeds the statistical errors present. This observation applies to constituents that are characterized by different effective magnetizations and/or by different gyromagnetic ratios. We have also shown that adjusted-$R^2$ values only provide limited guidance regarding the correct number of resonances to use when fitting spectra. However, careful inspection of the residuals of the fits of the spectra can provide important clues to identify the presence of additional resonances. In the absence of independent knowledge about the nature and number of resonances present in a material it is therefore advisable to pay close attention to the residuals of the fitted spectra.\\
Our investigations regarding the influence of noise present in ferromagnetic resonance spectra on the accuracy of the extracted parameters show very similar behavior as observed in the analysis of noiseless data. Not surprisingly noise acts to increase the variability of the extracted parameters and their error-margins. However, if the correct number of resonances are used to model the experimental spectra then extending the frequency range of the observations can significantly reduce the error-margins of both the extracted gyromagnetic ratio and the effective magnetization. We have also found that using a single resonance to fit spectra that clearly contain multiple resonances is surprisingly robust regarding noise present in the data. While there will certainly be exceptions to this, for the cases we investigated in this study, we found that the extracted values are in reasonable agreement with the weighted mean of the true values of the material. \\
However, our example for the extraction of inhomogeneous linewidth broadening and damping parameters from the frequency dependence of the resonance linewidth shows that the same cannot be said for these parameters. We see large deviations of those parameters from the true material parameters and their weighted means for any analysis that does not use the correct number of resonances present in the material. Even when the correct number of resonances is used for the analysis the results remain very sensitive to noise contamination of the data. It is therefore advisable to be careful when attempting to extract inhomogeneous linewidth broadening and damping parameters for materials that contain multiple resonances. As expected, minimizing the noise of the data while maximizing the frequency range over which data is collected is also advisable.

\section*{Acknowledgements}
\setlength{\parindent}{5ex}
We would like to thank Prof. John Xiao for helpful discussions and would like to thank him and Qorvo for providing the M-type hexaferrite sample. We would also like to thank Dr. Claudia Mewes for helpful discussions.
The authors would like to acknowledge funding for this research under the NASA Grant NASA CAN80NSSC18M0023.\par

\section*{Declaration of Competing Interest}
\setlength{\parindent}{5ex}
The authors declare that they have no known competing financial interests or personal relationships that could have appeared to influence the work reported in this paper.

\bibliography{References.bib}

\end{document}